\UseRawInputEncoding
\documentclass[amssymb,nobibnotes,superscriptaddress,notitlepage,twocolumn]{revtex4-2}

\usepackage{enumerate,amsmath}
\usepackage{graphicx}
\usepackage{color} 

\usepackage{braket}
\usepackage{qcircuit}
\usepackage{algpseudocode}
\usepackage{algorithm}
\usepackage{ulem}
\usepackage{here}

\usepackage{hyperref}
\hypersetup{
    colorlinks=true,
    linkcolor=blue,
    citecolor=magenta,
    filecolor=magenta,   
    urlcolor=cyan,
    }
\usepackage{url}

\begin{document}
\author{Yutaro Akahoshi}
\email{akahoshi.yutaro@fujitsu.com}
\thanks{These authors contributed equally to this work.}
\author{Riki Toshio}
\thanks{These authors contributed equally to this work.}
\author{Jun Fujisaki}
\author{Hirotaka Oshima}
\author{Shintaro Sato}
\affiliation{Quantum Laboratory, Fujitsu Research, Fujitsu Limited, \\4-1-1 Kawasaki, Kanagawa 211-8588, Japan}
\affiliation{Fujitsu Quantum Computing Joint Research Division, Center for Quantum Information and Quantum Biology, Osaka University, \\1-2 Machikaneyama, Toyonaka, Osaka, 565-8531, Japan}

\author{Keisuke Fujii}
\affiliation{Fujitsu Quantum Computing Joint Research Division, Center for Quantum Information and Quantum Biology, Osaka University, \\1-2 Machikaneyama, Toyonaka, Osaka, 565-8531, Japan}
\affiliation{Graduate School of Engineering Science, Osaka University, \\1-3 Machikaneyama, Toyonaka, Osaka, 560-8531, Japan}
\affiliation{Center for Quantum Information and Quantum Biology, Osaka University, 560-0043, Japan}
\affiliation{RIKEN Center for Quantum Computing (RQC), Wako Saitama 351-0198, Japan}

\title{Runtime reduction in lattice surgery utilizing time-like soft information}

\begin{abstract}
Runtime optimization of the quantum computing within a given computational resource is important to achieve practical quantum advantage. 
In this paper, we propose a runtime reduction protocol for the lattice surgery, which utilizes the soft information corresponding to the logical measurement error. 
Our proposal is a simple two-step protocol: operating the lattice surgery with the small number of syndrome measurement cycles, and reexecuting it with full syndrome measurement cycles in cases where the time-like soft information catches logical error symptoms. 
We firstly discuss basic features of the time-like complementary gap as the concrete example of the time-like soft information based on numerical results. Then, we show that our protocol surpasses the existing runtime reduction protocol called temporally encoded lattice surgery (TELS) for the most cases. 
In addition, we confirm that the combination of our protocol and the TELS protocol can reduce the runtime further, over 50\% in comparison to the naive serial execution of the lattice surgery. 
The proposed protocol in this paper can be applied to any quantum computing architecture based on the lattice surgery, and we expect that this will be one of the fundamental building blocks of runtime optimization to achieve practical scale quantum computing. 
\end{abstract}
\maketitle

\section{Introduction} \label{sec:intro}
Quantum computers are expected to bring exponential speedup over classical computers for various computational tasks, such as prime factoring~\cite{shor1999polynomial}, simulation of quantum many-body systems~\cite{abrams1999quantum,aspuru2005simulated}, and linear algebraic operations~\cite{harrow2009quantum}. 
Performing these complicated quantum algorithms requires the fault-tolerant quantum computing (FTQC), which suppresses quantum noise through the use of quantum error correction (QEC) codes in a scalable manner. 
One typical choice is the surface code-based FTQC architecture utilizing the lattice surgery~\cite{KITAEV20032,10.1063/1.1499754,PhysRevLett.98.190504,PhysRevA.86.032324,Horsman_2012}, which boasts a high noise threshold and good compatibility with the superconducting qubit devices. The recent surface code experiment clearly shows that a single logical qubit encoded by the surface code can achieve better fidelity than a physical qubit~\cite{google2024belowthesurfacecode}, indicating that we are now entering a new stage of the development of quantum computing, receiving benefits from the QEC on actual devices. 

However, the surface code-based architecture requires large qubit overhead due to its small encoding rate and the need for the magic state factories~\cite{gidney2021factor,yoshioka2022hunting,reiher2017elucidating,goings2022reliably}, motivating researchers exploring ways to reduce the resource overhead necessary for the practical quantum advantage. 
There are many studies ranging from the algorithm-level cost reduction~\cite{PRXQuantum.3.010318,PhysRevLett.129.030503,https://doi.org/10.48550/arxiv.2209.11322,PRXQuantum.4.020331}
to the architecture-level improvements like, converting QEC overhead to other forms (such as a sampling overhead) through quantum error mitigation techniques~\cite{suzuki2022quantum,piveteau2021error}, developing QEC codes that achieve better encoding rates~\cite{PRXQuantum.2.040101,Panteleev2021degeneratequantum,gidney2023yoked,hong2024longrangeenhanced,Bravyi_2024,goto2024manyhypercube,yoshida2024concatenate}, cost reduction of the magic state preparation~\cite{itogawa2025zeroleveldistill,gidney2024magicstatecultivationgrowing}. 
Recently, authors also propose the partially fault-tolerant quantum computing scheme, which is able to reduce the space-time overhead of the rotation gates by avoiding the magic state distillation and the decomposition to the Clifford + $T$~\cite{PRXQuantum.5.010337,PhysRevX.15.021057}. Some researches show the efficiency of this approach in the early-FTQC era~\cite{akahoshi2024compilationtrotterbasedtimeevolution,dangwal2025variationalquantumalgorithmsera}. 

One of the key techniques in such developments is the {\it soft information} of the QEC decoding~\cite{PhysRevA.89.022326,gidney2023yoked,PRXQuantum.5.010302}. The soft information is defined as the log-ratio of the probability of different logical sectors of decoding results; thus, it is naturally obtained by the maximum-likelihood decoder, like the tensor network decoder, but the high complexity of such decoders has prevented their applications in the practical setup. Recently, however, several ways to extract the soft information by utilizing efficient QEC decoding algorithm, such as the minimum-weight perfect matching (MWPM), are invented~\cite{gidney2023yoked,PRXQuantum.5.010302,meister2024efficientsoftoutputdecoderssurface,kishi2025toappear}, enabling us to estimate the reliability of the decoding result efficiently. Based on these inventions, several applications are discussed, such as the fault-tolerant postselection~\cite{PRXQuantum.5.010302,Smith_2024,gidney2024magicstatecultivationgrowing}, and the soft-decision decoding of the concatenated code~\cite{gidney2023yoked}. Since users can design many possible actions depending on the soft information extracted, it has broad possibilities for the resource optimization of the FTQC.

To the best of our knowledge, however, the  previous studies mainly focus on the soft information corresponding to logical Pauli errors {\it a.k.a} space-like logical error chains, so the discussion of the soft information on the time-like logical error chains during the lattice surgery, {\it a.k.a} logical measurement errors (in the following, we call it {\it time-like soft information}), has been limited. 
The time-like logical error commonly appears in the lattice surgery, so leveraging the corresponding soft information can bring another possibility for the resource optimization in the FTQC. 

In this paper, we propose a runtime reduction technique of the lattice surgery-based logical Pauli measurements by utilizing the time-like soft information. 
As mentioned above, we believe this work is the first to investigate the time-like soft information in details~\footnote{on preparation of the manuscript, we found a similar work~\cite{noah2025toappear}. Please see the note added in the last of Sec.~\ref{sec:summary}. }. 
As a concrete example of the time-like soft information, 
we first introduce the definition of the time-like complementary gap and discuss the basic property based on numerical simulations. 
We observe basic features similar to the space-like counterpart reported in the previous studies~\cite{gidney2023yoked}. 
Then, we discuss our proposed protocol to utilize the time-like soft information to the runtime reduction of the lattice surgery. 
Our protocol comprises two steps: (i) lattice surgery with the small number of syndrome measurement cycles, $L<d$, where $d$ is the code distance of the surface-code patch, together with the calculation of the time-like soft information, (ii) re-execution of the lattice surgery with the full syndrome measurement cycles, if the time-like soft information indicates the decoding result is not reliable. 
We show the efficiency of our protocol via a simple example of the logical $XX$ measurement between two surface code patches. 

We also compare our protocol with the existing runtime reduction protocol called {\it temporally encoded lattice surgery} (TELS)~\cite{PRXQuantum.3.010331}, which utilizes classical error correction codes to protect logical measurement results. We observe that our protocol surpasses it in most cases. 
Furthermore, we propose {\it soft information-accelerated TELS} (STELS), the high-level combination of our soft-information based protocol and the TELS protocol. 
As a result, using the STELS, we can go beyond the limitation achievable by sole use of them and achieve about 50\% reduction of the runtime over the naive lattice surgery. 
Since our protocol does not require any additional qubit overhead, 
the STELS pushes the limit of the space-time volume cost reduction discussed in Ref.~\cite{PRXQuantum.3.010331}. 
This proposal can be applied to any lattice surgery-based quantum computing architecture, especially suitable to the typical Litinski-style quantum computation~\cite{Litinski2019gameofsurfacecodes}. 
We expect that this will be one of the fundamental building blocks of runtime optimization to achieve practical scale quantum computing. 

This paper is organized as follows. 
In Sec.\ref{sec:gap}, we provide a definition of the time-like complementary gap considered in this paper, and discuss its behaviors based on numerical results. 
By utilizing those insights, in Sec.\ref{sec:accel}, we discuss how to reduce the runtime of the lattice surgery. 
We firstly summarize our protocol, then show its efficiency by a simple logical $XX$ measurement experiment. 
We then compare our protocol with the TELS protocol and propose the STELS protocol for further runtime reduction. 
Sec.\ref{sec:summary} is devoted to summary of our study and gives future directions. 

\section{Time-like complementary gap} \label{sec:gap}
In this section, we discuss basic behaviors of the time-like complementary gap in lattice surgery, which is a concrete example of the time-like soft information. 
We first give a definition of the complementary gap, following the previous study on the space-like counterpart~\cite{gidney2023yoked,PRXQuantum.5.010302}. 
Then, we show some numerical results and discuss its behaviors, comparing to the space-like counterpart~\cite{gidney2023yoked}. 

\begin{figure}
    \centering
    \includegraphics[width=0.5\linewidth, clip]{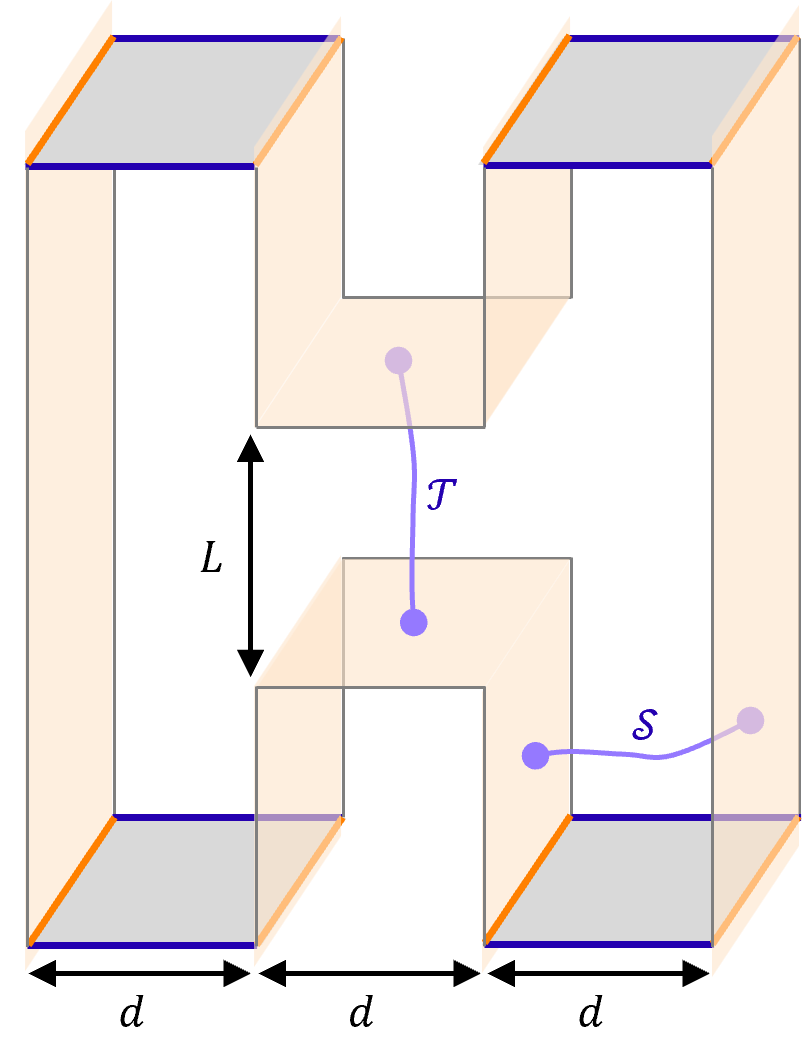}
    \caption{
    Space-time diagram of the logical $XX$ measurement by lattice surgery. 
    Time flows vertically in this figure. 
    We only show the $X$ boundaries of the space-time diagram (orange planes). 
    There are two possibilities of the logical errors during this operation: time-like logical errors (denoted by ${\mathcal T}$) and space-like ones (denoted by ${\mathcal S}$). In this paper, we consider the former. 
    The time length (= number of repeated syndrome measurements) during the merge operation is denoted as $L$ and is typically chosen as $L = d$ to sufficiently suppress the time-like logical error rate. 
    }
    \label{fig:XXmeas}
\end{figure}
Let us consider the logical $XX$ measurement operation between the two surface code patches with the code distance $d$, as shown in Fig.~\ref{fig:XXmeas}. 
During the operation, there are several possibilities that the logical errors happen. 
Among these, in this paper, we focus on the {\it time-like logical errors}, error chains connecting temporally separated boundaries, shown schematically as ${\mathcal T}$ in Fig.~\ref{fig:XXmeas}. 
This type of error chains results in the logical measurement error; in this example, the logical $XX$ measurement result will be flipped. 
To ensure the time-like logical errors are well suppressed comparable to the other types of logical errors, we have to perform the syndrome measurement at least $d$ times during the logical measurement (in this case $L = d$ in Fig.~\ref{fig:XXmeas}). 
This can be naturally generalized to the logical multi-Pauli measurement operations. 

Recently, several ways to quantify the {\it reliability} of the QEC decoding result have been proposed~\cite{PhysRevA.89.022326,gidney2023yoked,PRXQuantum.5.010302,meister2024efficientsoftoutputdecoderssurface}. In the following, we call such a reliability index {\it soft information}. 
These previous studies mainly focus on the {\it space-like logical errors}, error chains connecting spatially separated boundaries (like ${\mathcal S}$ in Fig.~\ref{fig:XXmeas}), and discuss how to detect the symptoms of such logical errors by using the soft information. We extend these concepts to the time-like logical error and utilize it for the runtime optimization of the lattice surgery. 

Following Ref.~\cite{PRXQuantum.5.010302}, we define the signed time-like complementary gap as 
\begin{equation}
    \Delta_{\mathcal T}(s) \equiv \bar w_{l_{\rm wrong}} (s) - \bar w_{l_{\rm correct}} (s),  
\end{equation}
where $\bar w_{l}(s)$ is the log-likelihood weight of the correction $l$ for the given syndrome $s$ for the time-like logical error sector. 
The correction $l_{\rm wrong}$ ($l_{\rm correct}$) together with actual errors yields the time-like logical error chain ${\mathcal T}$ (logical identity operator ${\mathcal I}$).
In this study, we use the minimum-weight perfect matching (MWPM) decoder, which outputs the most likely (smallest weight) error pattern. 
Therefore, the negative $\Delta_{\mathcal T}$ indicates that the time-like logical error is introduced by the correction. 
In practice, however, we can only calculate the absolute value $g = |\Delta_{\mathcal T}|$ since which correction is correct is unknown. We simply call this absolute value {\it time-like complementary gap} or {\it gap} in the following discussion. 
Once we calculate the time-like complementary gap, we can get the reliability information from it: 
For example, if we obtain the large gap, $g \gg 1$, the probability that the weight of the correct sector is far smaller than that of the wrong sector, is exponentially large, indicating that the decoding result is highly likely to be the correct sector. 
Therefore, the decoding result is highly reliable. 
On the other hand, if the gap is very small, $g \approx 0$, both the correct and wrong sectors have almost the same weight, therefore both of the decoding results happen evenly. It means that the decoding result is not reliable. 
In this way, we can detect the symptom of the logical measurement errors during the lattice surgery via the time-like gap. 

\begin{figure}
    \centering
    \includegraphics[width=80mm,clip]{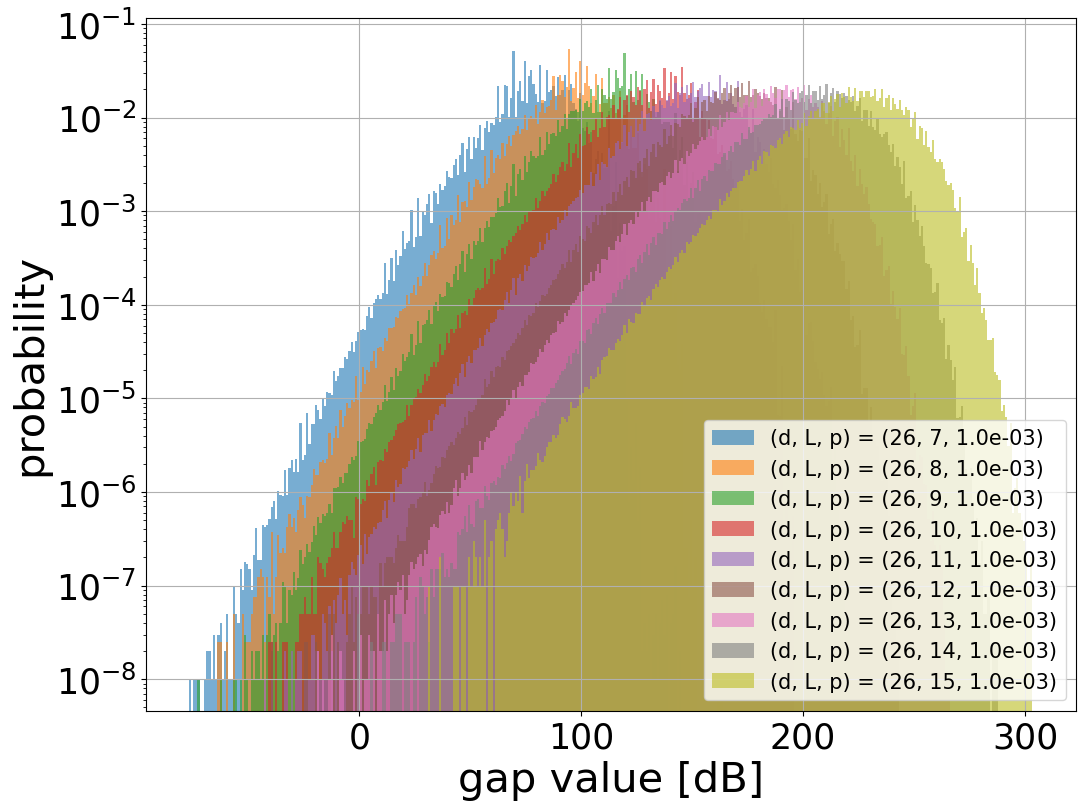}
    \caption{Distribution of the signed time-like gap obtained from the large stability experiment with $d = 26, L = \{ 7,8,9,10,11,12,13,14,15 \}$. 
    }
    \label{fig:gap_dist_larged_Ldep}
\end{figure}

We numerically calculate the distribution of the signed time-like complementary gap. 
We use the open-source python library Stim~\cite{Gidney_2021} and PyMatching~\cite{pymatchingv2}, and perform numerical simulation of the lattice surgery operation shown in Fig.~\ref{fig:XXmeas} (in the following, we call this type of experiment ``H-shape experiment" after its space-time diagram). The details of the numerical simulation are given in Appendix~\ref{appx:numericalsim}.  
Figure~\ref{fig:gap_dist} (blue) shows the distribution of $\Delta_{\mathcal T}$ with $p = 10^{-3}, d = L = 9$. 
We obtain a similar distribution as the space-like complementary gap, reported in Ref.~\cite{gidney2023yoked}. 
We observe a sudden cutoff behavior at the large gap region around 210 dB, but we think it mainly comes from the finite-volume effect (see the larger size numerical simulation discussed later, Fig~\ref{fig:gap_dist_larged_Ldep}). 
Since the distribution of the large gap region is not important in the following discussion, we do not dive into this part deeper. 

Next, we simulate another type of experiments, so-called {\it stability experiment}~\cite{Gidney2022stability}. 
This experiment mimics the structure of the routing region during the lattice surgery. 
Since our main focus is the time-like logical errors, which happen in the routing region, we expect that the stability experiment captures most of the important features of the time-like gap. 
We show the details of the patch configuration for the stability experiment in Appendix~\ref{appx:numericalsim}. 
As shown in Fig.~\ref{fig:gap_dist} (orange), if we choose the similar patch size as the H-shape experiment, we confirm that the distribution is almost the same except for the large value region. 
We observe the isolated largest gap value around 210 dB in the stability experiment distribution. 
This value is equal to the weight sum of the shortest path between two time-like boundaries, and the second largest gap value around 195 dB comes from the single measurement error configuration somewhere in the syndrome extraction. 
Since there are no error events realizing the gap value between these two largest values in the stability experiment, there is a chasm around 200 dB. 
On the other hand, in the H-shape experiment, there are some marginal gap events between them via the leg part of the space-time diagram, so the distribution at the large value does not show any chasm structure. 
Note that these behaviors at the large gap region is also significant here due to the finite-volume effect (as seen later in Fig.~\ref{fig:gap_dist_larged_Ldep}, such behaviors become less noticeable with the large-sized patch). 
In the following, we focus on the stability experiment results since it enabled us to perform large size simulation necessary for our discussion.

\begin{figure}
    \centering
    \includegraphics[width=80mm,clip]{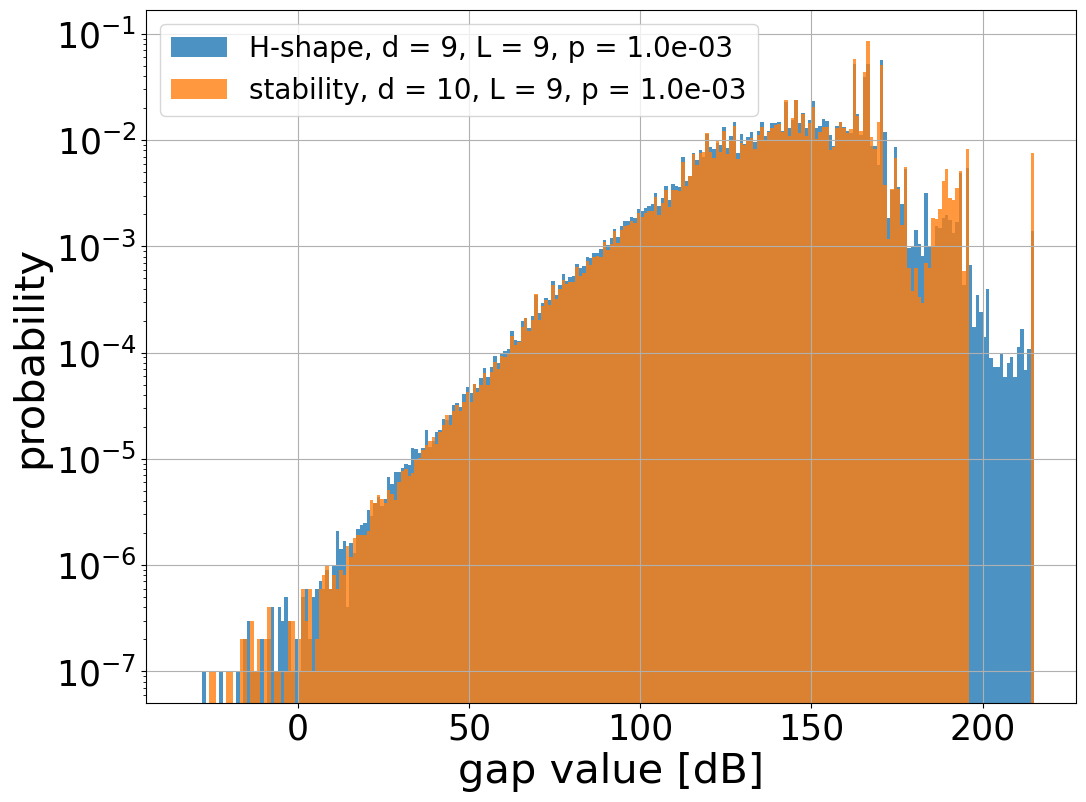}
    \caption{Distribution of $\Delta_{\mathcal T}$. Blue and orange histograms are obtained from the logical operation described in Fig.~\ref{fig:XXmeas} with $d = L = 9$ and the stability experiment setup with $d = 10, L = 9$, respectively. }
    \label{fig:gap_dist}
\end{figure}

\begin{figure}
    \centering
    \includegraphics[width=80mm,clip]{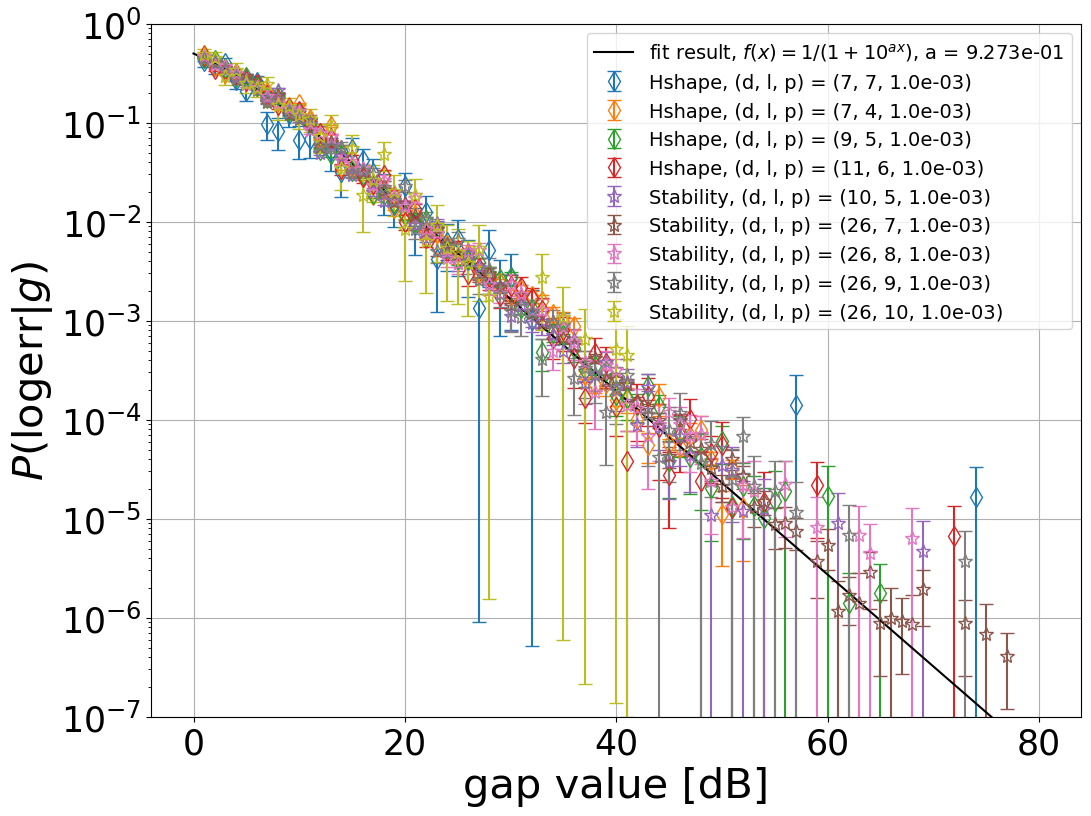}
    \caption{Conditional logical error probability by the time-like gap value. We use several distributions obtained by different $(d, L)$ values, as well as changing the experimental setup (H-shape or stability). The solid black line shows the global fit result using the single-parameter fit function $f(g) = \frac{1}{1 + 10^{ag}}$.}
    \label{fig:cond_prob}
\end{figure}

We also check the behavior of the logical error probability conditioned by the time-like gap $g$, $P({\mathcal T}|g)$. 
This quantity is important for later discussions, and also the previous study has reported its interesting universal behavior~\cite{gidney2023yoked}. 
Figure ~\ref{fig:cond_prob} shows $P({\mathcal T}|g)$ obtained by several different sets of $(d, L)$, and both the H-shape and the stability setups. 
We observe all of them have the same universal dependency on the time-like gap value, fitted as 
\begin{equation} \label{eq:cont_prob_scaling}
    P({\mathcal T}|g) = \frac{1}{1 + e^{a g }}
\end{equation}
with $a = 0.9273$ (1/B) (fitting result is also shown in Fig.~\ref{fig:cond_prob}). 
As mentioned, this kind of universal behavior has been observed in the previous space-like gap analysis~\cite{gidney2023yoked}, 
and our result is very similar value to the previous reported value $a = 0.9$ (1/B). 
This suggests that the gap behavior is universal regardless of whether it is space-like or time-like. 
Naively, if we see the lattice surgery operations as the 3D topological object, distinction between time-like and space-like strings seems to be ambiguous, so the universal behavior seems to be natural. 
In practice, on the other hand, the dominant error mechanism that brings the time-like and space-like logical errors is different, so the scaling factor $a$ can different in general. 
We do not discuss this universal behavior further because it is out of our main interest, but we think that it is a theoretically interesting topic that should be addressed in future. 

To reduce the runtime of the lattice surgery, we decrease the value $L$ from $d$ in the later discussion. 
To obtain necessary information, we calculate the distributions of the time-like gap for several $L$ by the stability experiment. 
Figure~\ref{fig:gap_dist_larged_Ldep} shows these distributions with $p = 10^{-3}$ and $d = 26$, which is chosen for the later discussion based on $d = 25$. 
This choice is inspired by the study on resource estimation~\cite{yoshioka2022hunting}, in which the practical quantum advantage is expected with $p = 10^{-3}$ and $d = 25$. 
We observe clear dependence on $L$, which is naturally expected from the space-like counterpart~\cite{gidney2023yoked}. 
We also confirmed that the time-like logical error rate exponentially decreases as $\propto p^{\lfloor \frac{L+1}{2}\rfloor}$ (see Appendix~\ref{appx:lerfit}). 

\section{Accelerating logical Pauli measurements by time-like soft information} \label{sec:accel}
\begin{figure*}
    \centering
    \includegraphics[width=0.6\linewidth,clip]{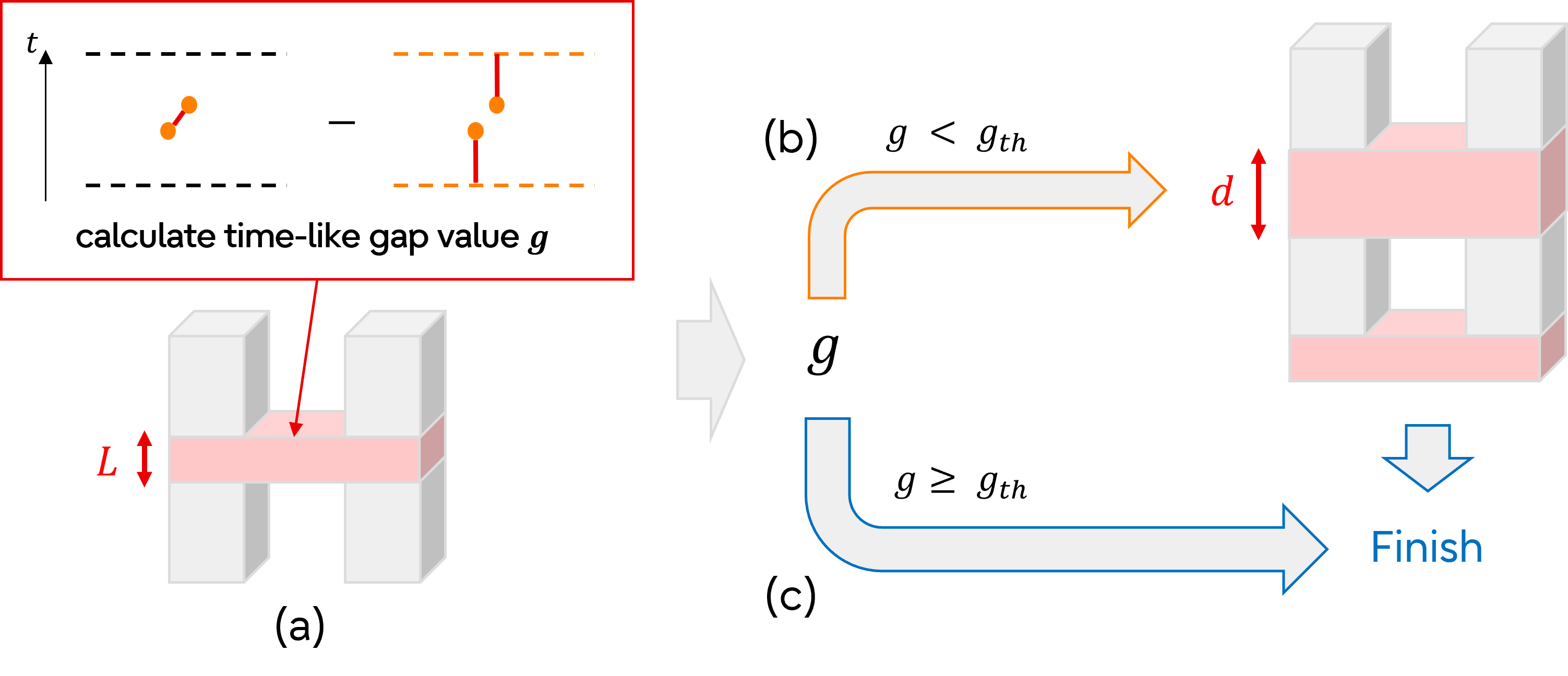}
    \caption{
    Overview of the proposed runtime reduction protocol.
    (a) Performing the lattice surgery operation with a reduced syndrome measurement cycle, $L$. The time-like gap $g$ is calculated here. 
    (b) If $g$ is smaller than the threshold value $g_{\rm th}$, we perform the lattice surgery operation with a full syndrome measurement cycle, $d$, to ensure the resultant logical measurement value. 
    (c) If $g$ is larger or equal to $g_{\rm th}$, we can move on to the next operation. 
    The optimal value $g_{\rm th}$ is determined by the condition that the logical error rate of the entire protocol keeps the same value of the original lattice surgery operation with $d$ cycle syndrome measurements (details are given in Appendix~\ref{appx:g_thdet}). 
    }
    \label{fig:overview}    
\end{figure*}
\begin{figure}
    \centering
    \includegraphics[width=85mm,clip]{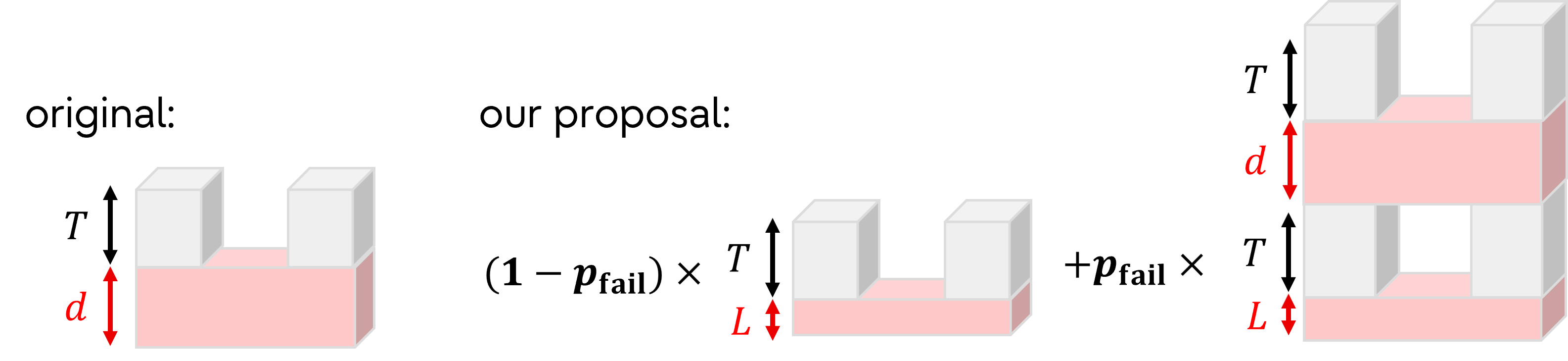}
    \caption{
    Schematic comparison of the average runtime between the naive lattice surgery (original) and our proposed protocol (our proposal). 
    }
    \label{fig:averuntime}    
\end{figure}
Given the definition of the time-like gap and its basic behavior, in this section, we discuss how to utilize it to the runtime reduction in the lattice surgery. 
We focus on the logical multi-Pauli measurements, which is the fundamental building block of the typical surface code-based FTQC~\cite{Litinski2019gameofsurfacecodes}. 
Firstly, we summarize our proposed protocol. 
Then, we apply it to the simplest lattice surgery operation, a two-Pauli measurement (H-shape experiment), and estimate its performance. 
After that, we compare our protocol to the previously proposed runtime reduction technique called {\it temporally encoded lattice surgery (TELS)}. 
Finally, we propose the combination of these two techniques, soft-information-accelerated TELS (STELS), and show that the STELS can achieve the best performance. 

\subsection{Proposed protocol} \label{sec:protocol}
Overview of our protocol is given in Fig.~\ref{fig:overview}. 
Our protocol is a simple two-step protocol: 
(i) performing a logical measurement with smaller syndrome measurement rounds, $L < d$, and calculating the time-like gap $g$ between two closed boundaries on the top and bottom of the merged region (Fig.~\ref{fig:overview} (a)), 
(ii) performing a remeasuement with full syndrome measurement rounds, $L=d$, if the obtained gap $g$ is smaller than a threshold value $g_{\rm th}$ (Fig.~\ref{fig:overview} (b)), otherwise the measurement is finished (Fig.~\ref{fig:overview} (c)). 
By using this protocol, if the remeasurement probability $p_{\rm fail}$ is sufficiently small, we can reduce the average runtime from $d + T$ to $(1-p_{\rm fail}) (L + T) + p_{\rm fail} (L+T+d+T) \approx L + T$ in units of syndrome measurement round, where $T$ is the reaction time, measured in units of syndrome measurement round. The reaction time $T$ includes the decoding time for obtaining the logical measurement result and the time-like gap and the time for deciding a next logical operation (see Fig.~\ref{fig:averuntime}). 

The threshold value $g_{\rm th}$ is determined by the condition that the logical error rate of the entire protocol should be the same as the naive protocol (not using the time-like gap) with $L = d$. 
The entire logical error rate of our protocol is given as 
\begin{eqnarray} \label{eq:total_ler}
    && P_{>g_{\rm th}}(g_{\rm th}) + p_{\rm fail}(g_{\rm th}) P_{\rm full} \\
    &=& \int_{g_{\rm th}}^{\infty} dg P({\mathcal T}|g) P(g) + P_{\rm full} \int_{0}^{g_{\rm th}} dg P_L(g), \nonumber
\end{eqnarray}
where $P_L(g)$ is the gap distribution with the reduced syndrome measurements $L$ and $P_{\rm full}$ is the logical error rate of the naive protocol with $L = d$. 
The first term, $P_{>g_{\rm th}}(g_{\rm th})$, is the thresholded logical error rate coming from the success branch (Fig.~\ref{fig:overview} (c)), and the second term is the contribution from the failure branch (Fig.~\ref{fig:overview} (b)). 
The optimal threshold value $g^*_{\rm th}$ should satisfies
\begin{equation}
    P_{>g_{\rm th}}(g^*_{\rm th}) + p_{\rm fail}(g^*_{\rm th}) P_{\rm full} = P_{\rm full}.
\end{equation}
The failure rate is immediately obtained as $p_{\rm fail}(g^*_{\rm th})$.
In the following, we determine $g^*_{\rm th}$ from the numerically obtained time-like gap distribution, shown in Fig.~\ref{fig:gap_dist_larged_Ldep}. 
For example, for a simple logical two-Pauli measurement discussed in the next section, 
the $g^*_{\rm th}$ is about 120-140 dB depending on $L$. The detailed process and actual values of $g^*_{\rm th}$ can be found in Appendix~\ref{appx:g_thdet}.

\subsection{Simple example: two-qubit Pauli measurement} \label{sec:simpleex}

\begin{figure}
    \centering
    \includegraphics[width=0.9\linewidth,clip]{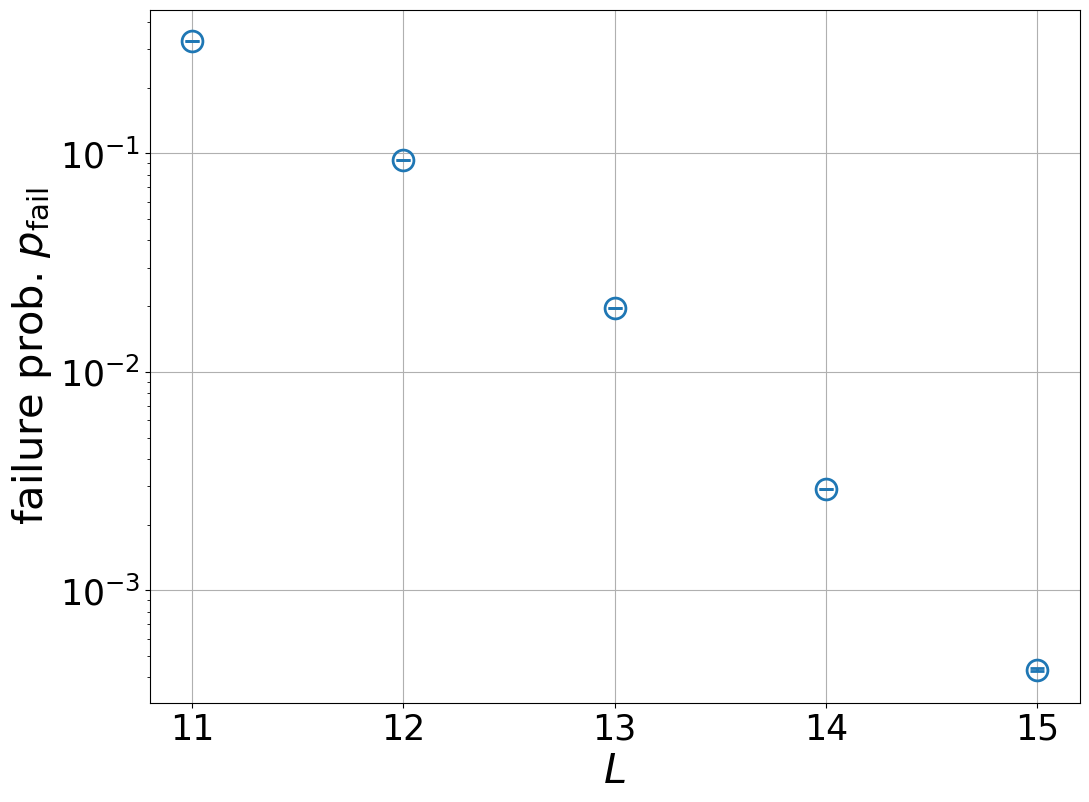}
    \caption{Failure probability $p_{\rm fail}$ estimated by different $L$.}
    \label{fig:pfail_simpleXX}    
\end{figure}

\begin{figure}
    \centering
    \includegraphics[width=0.9\linewidth,clip]{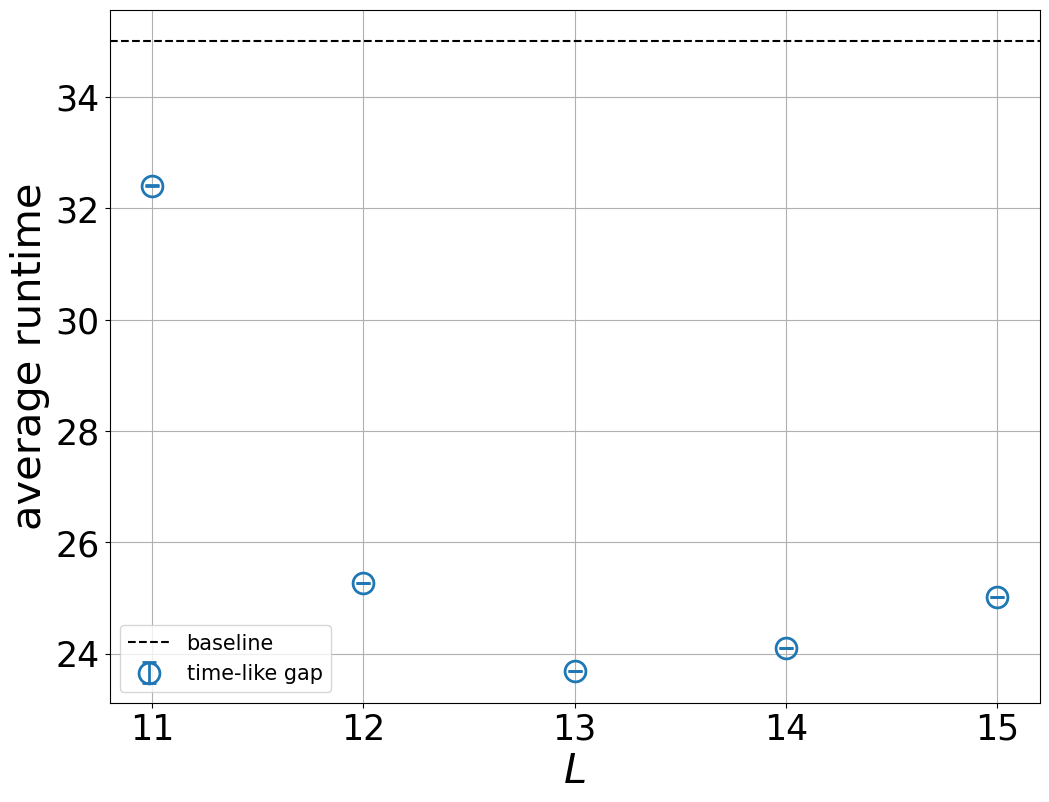}
    \caption{Average runtime result. $L = 13$ is the best parameter in this setup. The dashed line indicates the runtime of the naive protocol, namely $d+T = 35$.}
    \label{fig:averuntime_simpleXX}
\end{figure}

Firstly, we justify the advantage of our protocol for the simplest case, the logical two-Pauli measurement shown in Fig.~\ref{fig:XXmeas}. 
We assume $p = 10^{-3}$, $d = 25$ in this estimation and use numerical results of the stability experiment with $d = 26$. 
We set $T = 10$ since some resource estimation studies~\cite{yoshioka2022hunting,gidney2021factor} assume that the single syndrome measurement round and the reaction time take 1 $\mu s$ and 10 $\mu s$, respectively. 
In Fig.~\ref{fig:pfail_simpleXX}, we show the value of $p_{\rm fail}$ obtained with several $L$, ranged from 11 to 15. The details of their determination can be found in Appendix~\ref{appx:g_thdet}. 
We observe that the $p_{\rm fail}$ monotonically decreases with increasing $L$, which is naturally understood: For the small $L$, the logical error chains are easily introduced, therefore we have to make the criteria of the post-selection tighter to ensure the post-selected logical error rate is the same as the original one with $L = d$. 
Figure~\ref{fig:averuntime_simpleXX} shows the estimated average runtime. 
As a result, the average runtime of $L = 11, 12$ is worse than others due to the large $p_{\rm fail}$ that makes the remeasurement overhead not negligible. 
In this simple example, $L = 13$ is optimal and we can achieve about 32\% runtime reduction on average. 


\subsection{Comparison to temporally encoded lattice surgery (TELS)} \label{sec:tels}
Next, we compare our protocol to the existing runtime reduction protocol called TELS~\cite{PRXQuantum.3.010331}. 
The basic idea behind TELS is protecting logical measurement results by some classical error correction codes (these codes are called {\it measurement code} in the following). 
Let us consider the sequence of the $k$ different logical Pauli measurements for ${P_1, P_2, ..., P_k}$, which commute with each other. 
In the simplest TELS protocol, which utilizes the $[k+1, k, 2]$ measurement code, we add a measurement of the product of all $k$ Pauli operators, $P_{k+1} = P_1 P_2 \cdots P_k$. 
The measurement result of the additional Pauli operator works as the error syndrome since the measurement results $m_i \quad (i = 1,...,k+1)$ should satisfy $m_{k+1} = \sum_{i=1}^{k} m_i$. 
Therefore, we can suppress the logical measurement error in this sequence as $O(p_{\rm meas}^2)$, where $p_{\rm meas}$ is the logical error probability of each measurement. 
By using this, we can reduce the time length $L$ of each measurement while keeping the logical error probability. 
From the asymptotic behavior, we expect that $L = d/2$ is optimal in principle, but in practice the optimal $L$ is slightly larger than $L = d/2$ due to the constant factors. 
In general, we can choose how to treat the erroneous events: error correction mode or error detection mode. In the original paper~\cite{PRXQuantum.3.010331}, however, the authors observed that the error detection mode is always superior to the error correction mode. 
Following this observation, we basically consider the error detection mode in the performance estimation of the TELS protocol. 

\begin{figure}
    \centering
    \includegraphics[width=\linewidth,clip]{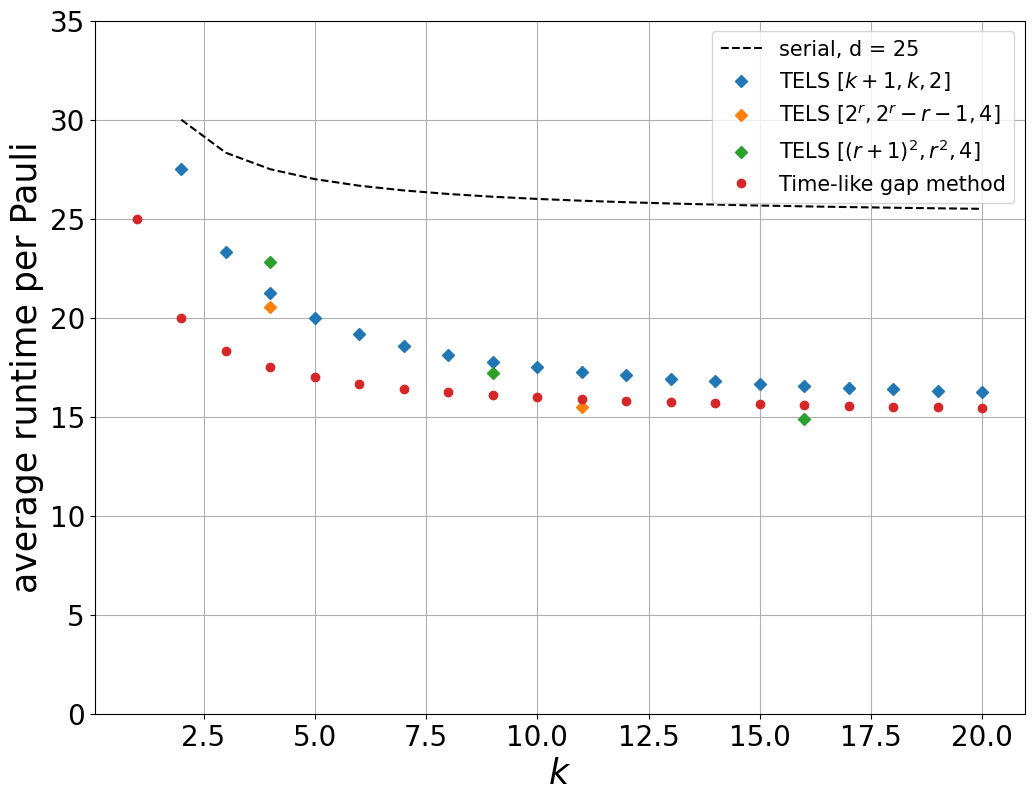}
    \caption{Comparison between our soft-infomation based time reduction and TELS.}
    \label{fig:comp_tels}
\end{figure}

We compare the runtime reduction performance of our protocol and TELS. 
We assume $d = 25$, $p = 10^{-3}$, and $N=10$ logical qubits, and calculate average runtime per Pauli measurement for several $k$. 
In the TELS performance estimation, we consider several measurement codes discussed in the original paper~\cite{PRXQuantum.3.010331}: error detection code ($[k+1, k, 2]$), extended Hamming code ($[2^r, 2^r-r-1,4]$), and concatenation of the $[k+1, k, 2]$ code ($[(r+1)^2, r^2, 4]$). 
The details of this analysis can be found in Appendix~\ref{appx:comp_details}. 
The results are summarized in Fig.~\ref{fig:comp_tels}.
As seen in Fig.~\ref{fig:comp_tels}, for most of $k$, our proposal is superior to the TELS protocol. 

The first advantage of our protocol is that there is no need for redundant measurements. 
Thanks to this, in the small $k$ region, where the overhead of redundant measurements is significant, our protocol has much better performance. 
Furthermore, applicability to $k = 1$ is another advantage of our protocol. 
Naively speaking, the TELS protocol is also applicable to $k=1$, but in this case the runtime does not decrease due to the existence of the redundant measurement. 
For example, let us consider the $[2,1,2]$ measurement code. 
The runtime is $2L + T$ when we ignore the remeasurement overhead, but even if the ideal limit $L = d/2$ is achieved, it is just the same as the naive runtime, $d + T$. 

We observe that the TELS protocol using large code distance measurement codes outperforms our protocol at $k = 11, 16$. 
This indicates that the TELS protocol can be better if the appropriate measurement code is found for the target $k$. 
To maximize the advantage of the TELS protocol, we have to, for example, split the sequence of Pauli measurements into layers with a certain number of Pauli measurements where the high performance measurement code is applicable. 
On the other hand, our protocol can be applicable to any $k$ and consistently delivers high performance regardless of the value of $k$. 
We expect that the ability to achieve high performance with simple two-step protocol can also be an advantage from the perspective of a device's control unit.

\subsection{STELS: Soft-information-accelerated temporally encoded lattice surgery} \label{sec:stels} 
\begin{figure*}
    \centering
    \includegraphics[width=0.7\linewidth,clip]{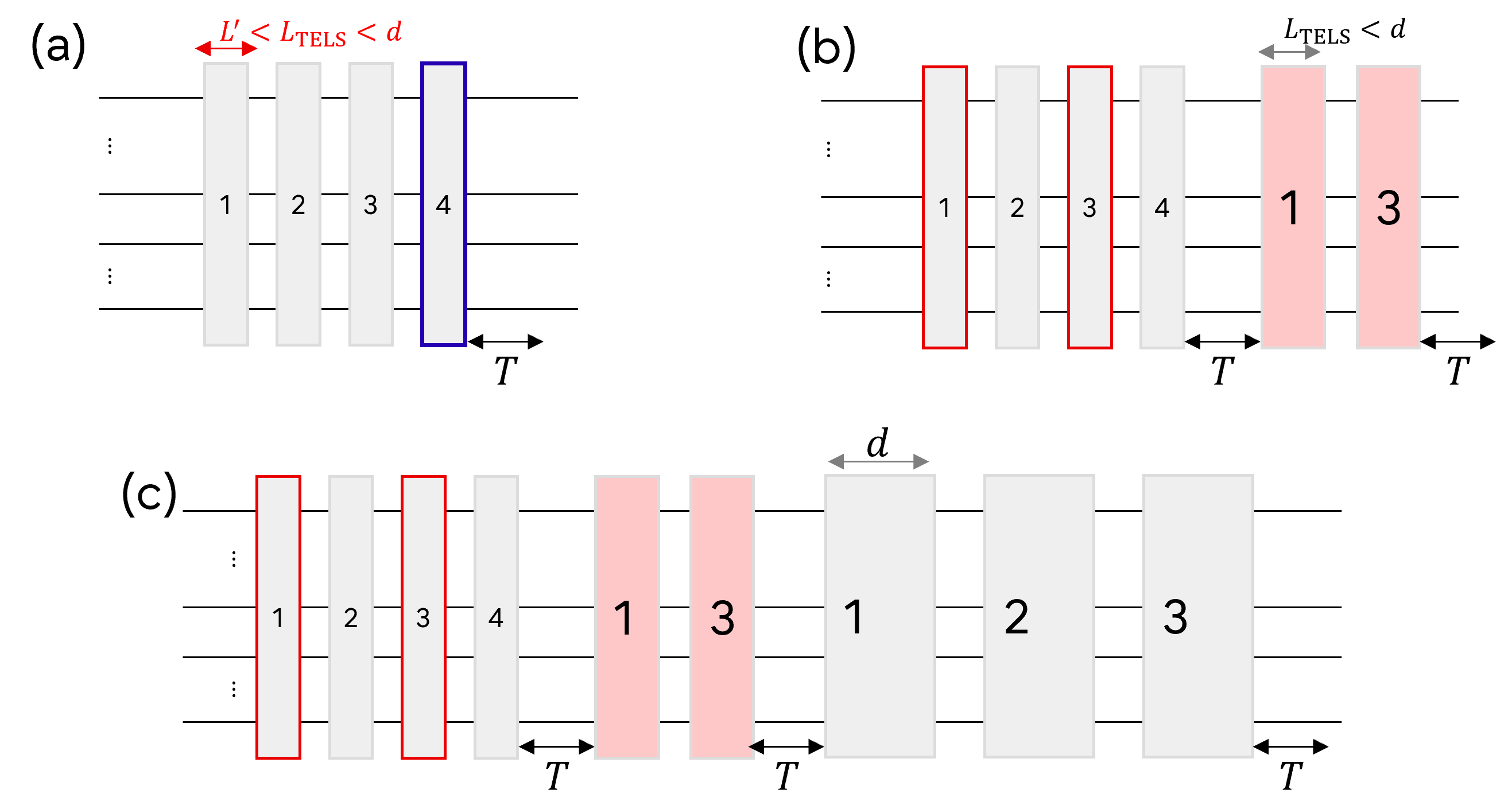}
    \caption{
    Overview of the soft-information-enhanced TELS (STELS) protocol. This figure shows $k = 3$, $[k+1, k, 2]$ measurement code example. 
    (a) Measuring all Pauli operators including the redundancy one (blue encircled) with the time interval $L'$, which is smaller value than that of the TELS. For each measurement, we calculate the time-like complementary gap. 
    We assume that we can decode syndromes from a logical Pauli measurement during performing the next Pauli measurement; thus, the reaction time $T$ only appears after the last measurement. This assumption also holds in the following steps. (b) If some measurements has small gap value than $g_{\rm th}$, they are remeasured (in this example, \#1 and \#3 are remeasured). These remeasurements are done with the optimal time interval $L_{\rm TELS} > L'$ in the TELS protocol. 
    (c) By using the measurement values obtained in (a), (b), we check the parity constraints between them. If some parity checks are failed, all the Pauli operators are measured once again with the full time interval $L = d$. 
    }
    \label{fig:stels_overview}
\end{figure*}

Finally, we show that the combination of the two protocols enables us to reduce the runtime further. 
The two protocols utilize different information to reduce the runtime of the LS: time-like soft information from the decoding (our protocol) and the parity information from the encoded measurement results (TELS). Since those information can be obtained at the same time, we can combine these two protocols. 
We call this combinatorial protocol {\it soft-information-accelerated TELS (STELS)}. 
The schematic overview of the STELS is shown in Fig.~\ref{fig:stels_overview}. 

The STELS can be divided into two steps: (i) Runtime reduction of the first measurement layer using our protocol, and (ii) Error detection using the TELS to reproduce original measurement precision. 
The first step measures all Pauli operators including redundancies with $L' < d$ syndrome cycles and calculate the time-like gap (Fig.~\ref{fig:stels_overview} (a)). 
By using our protocol for each Pauli measurement, we can choose smaller $L'$ than $L_{\rm TELS}$, where $L_{\rm TELS} < d$ is the optimal number of syndrome cycles of the single use of the TELS protocol, while keeping its precision (Fig.~\ref{fig:stels_overview} (b)). 
Note that during the first step, all we have to do is to keep the logical error rate as it is with $L_{\rm TELS}$ (no need to keep the original measurement precision with $L = d$), since logical measurement errors happening in this step can be detected in the next step. See Appendix~\ref{appx:stels_gth_determination} for details. 
After collecting all measurement values, the second step is to check the parity of the redundancy measurements. If there are some error symptoms, we remeasure all $k$ Pauli operators with the full time interval $L = d$ (Fig.~\ref{fig:stels_overview} (c)). 

\begin{figure}
    \centering
    \includegraphics[width=\linewidth,clip]{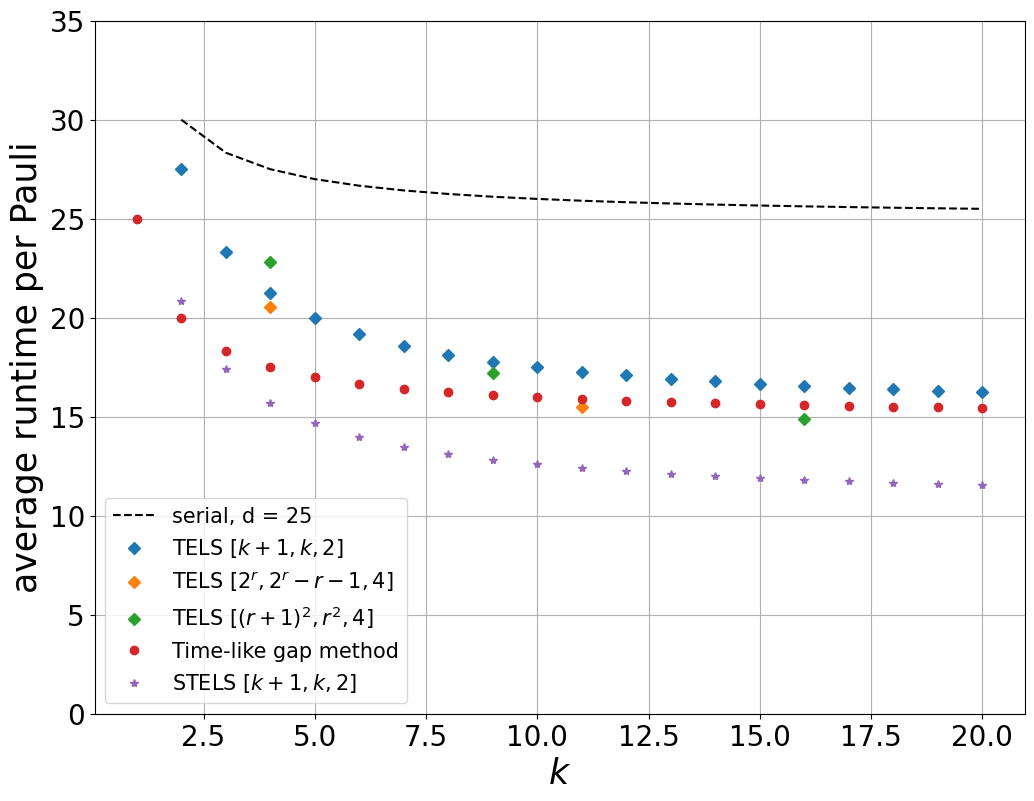}
    \caption{Performance of the STELS protocol using a $[k+1, k, 2]$ measurement code.}
    \label{fig:stels_performance}
\end{figure}

We investigate the performance of the STELS protocol with the same setup as discussed in Sect.~\ref{sec:tels}, namely $d = 25, p=10^{-3}, N=10$. 
As the measurement code, we employ the simplest $[k+1, k ,2]$ code because of the wide applicability to the value of $k$. 
Figure \ref{fig:stels_performance} shows the average runtime per Pauli measurement with given $k$. 
After the optimization of the parameters, we obtain $L_{\rm TELS} = 15$ and $L' = 10$. 
Thanks to the time-like gap, we can further reduce the number of syndrome cycles from 15 to 10 while keeping the entire logical error rate. 
As a result, the STELS surpasses all other protocols except for $k = 2$. 
The worse performance at $k = 2$ is mainly attributed to the redundant measurement overhead from the TELS part. 
At $k = 2$, the single usage of the time-like gap method has a reduced runtime $2L + T = 2 \times 14 + 10 = 38$ (ignoring the remeasurement). 
On the other hand, the STELS takes $3\times L' + T = 3 \times 10 + 10 = 40$ (ignoring the remeasurement), so it is natural that the STELS is worse than the single usage of the time-like gap method at $k = 2$. 
For large $k$, the overhead gradually becomes negligible and the STELS shows significantly better performance than others, over 50\% runtime reduction is achieved in comparison to the naive protocol. 

Finally, we comment on the use of the STELS protocol in the error correction mode. 
Same as the TELS protocol, the STELS protocol supports the error correction mode. 
We can mitigate the performance gap between the error correction mode and the error detection mode by utilizing time-like complementary gap. 
Let us consider the STELS using the code distance $D$ measurement code and the number of syndrome cycles $L$ for each measurement. 
In this setup, the STELS protocol can be interpreted as the concatenation of the classical codes with the code distance $L$ (lattice surgery with the time interval $L$) and $D$ (measurement code). 
In general, we can decode such a concatenated code by using two types of the decoding strategies: the hard-decision decoding and the soft-decision decoding. To achieve the theoretically expected code distance of the concatenated code, we should use the soft-decision decoding~\cite{PhysRevA.74.052333}. 
The original TELS in the error correction mode is based on the hard-decision decoding, but in the STELS protocol, we can use the time-like complementary gap for the soft-decision decoding. 
As a result, the logical error rate asymptotically achieve $O(p^{\frac{LD+1}{2}})=O(P^D)$ in principle, the same order as the error detection mode. It means that we can overcome the main obstacle of the TELS in the error correction mode, the restricted performance due to the hard-decision decoding. 
This is another advantage of the STELS. 

\section{Conclusion and outlook} \label{sec:summary}
In this paper, we propose the runtime reduction protocol of the lattice surgery by utilizing the time-like soft information. 
In our proposed protocol, we reduce the number of syndrome measurement cycles during the lattice surgery and check the time-like gap to determine whether the operation is completed without time-like logical errors or not. 
If the time-like gap indicates the error symptoms, we retry the lattice surgery operation with the longer syndrome cycles to suppress logical errors. As a result, the average runtime decreases when the retry probability is sufficiently small. 
We first investigate the basic features of the time-like complementary gap, 
observing that it is similar to the space-like counterpart studied before. 
Then, based on the numerical results, we confirm that our protocol shows better performance than the previously proposed TELS protocol in the most cases. 
Moreover, we show that we can achieve further runtime reduction over the sole use of the TELS and our protocol by combining these two protocols, STELS. 

Finally, we comment on how our protocol is affected by the compilation scheme of FTQC. 
Our proposal has good compatibility with the sequential compilation of the FTQC~\cite{Litinski2019gameofsurfacecodes}, since the compiled circuit is the sequence of the multi-Pauli non-Clifford gates, which consist of the multi-Pauli measurements. 
In our analysis, we assume a simple situation, where each Pauli measurements are performed sequentially and the basis change overhead during the lattice surgery is omitted (see Sec.~\ref{appx:comp_details}). 
In practice, however, the basis change overhead is significant in some logical patch arrangement, as well as the parallel Pauli measurement can be done if the routing region does not overlap. 
Furthermore, the layering of the sequence of the Pauli measurement suitable for the STELS protocol is also important. 
Therefore, the efficient compilation method to maximize the effect of the STELS protocol is needed. The runtime estimation of some practical computational tasks using such a compilation method is also important. 
In cases where the direct Clifford + $T$ compilation is used, on the other hand, we expect that the merit of our protocol is limited. 
Although the single $T$ gate via the gate teleportation circuit can be accelerated by our protocol, runtime reduction for other gates, such as the logical CNOT gate, can be achieved only when $d > L+T$ is satisfied, because the reaction time overhead can be hidden due to the Pauli frame tracking. We give a short explanation of this limitation in Appendix~\ref{appx:cnot_circuit}. 
There are several works on the optimization of the direct Clifford + $T$ compilation~\cite{PRXQuantum.3.020342,hamada2024efficient,tan2024sat}, so investigating whether combining such optimization techniques and our protocol provides further runtime reduction is one possible future direction. 

\vspace{5mm}
NOTE ADDED: Upon preparation of the manuscript, we found a similar work on the QEC25 conference archive~\cite{noah2025toappear}. 
They also investigated the runtime reduction of the lattice surgery operation (the H-shape space-time diagram discussed in this paper) via complementary gap and found that they can reduce the average number of the syndrome measurements to slightly larger than $d/2$ while keeping the whole logical error rate remaining. 
Our work distinguishes itself from theirs by conducting a comparative analysis with TELS, and further illustrates that superior performance can be attained through STELS, the integration with TELS. 

\begin{acknowledgments}
K.F. is supported by MEXT Quantum Leap Flagship Program (MEXT Q-LEAP) Grant No. JPMXS0120319794, 
JST COI-NEXT Grant No. JPMJPF2014, JST Moonshot R\&D Grant No. JPMJMS2061, and JST CREST Grant No. JPMJCR24I3.
\end{acknowledgments}

\appendix
\section{Simulation details} \label{appx:numericalsim}
In this appendix, we discuss details of our numerical simulations, namely the noise model, qubit configurations of the surface code lattice surgery experiment and the stability experiment discussed in the main text. 
\subsection{Noise model}
In our numerical simulation, we employ the circuit-level noise model. 
We assume that the native physical gates are Hadamard and CNOT, initialization to $\ket 0$, and measurement in the $Z$ basis; 
thus, we employ the depth-8 syndrome measurement circuit.
Noisy initialization and measurement flip to an orthogonal state with a probability $p$, and noisy Hadamard and CNOT gates are modeled by ideal operations followed by the depolarizing noise channels, 
\begin{equation} \label{eq:single_depolarizing}
  {\mathcal E}_{\rm single}(\rho) = (1-p) \rho + \frac{p}{3} \left( X \rho X + Y \rho Y + Z \rho Z \right), 
\end{equation}
and
\begin{eqnarray} \label{eq:double_depolarizing}
  && {\mathcal E}_{\rm double}(\rho) = \left(1-\frac{16}{15}p \right) \rho 
    + \frac{p}{15}
\sum _{E \in \{ I,X,Y,Z \}^{\otimes 2} } E \rho E , \nonumber \\
\end{eqnarray}
respectively. 
Noisy identity gates are inserted wherever physical qubits are idle. 
We assume that all errors occur with a common probability $p = 10^{-3}$. 

\subsection{Qubit configuration of the logical $XX$ measurement experiment}
\begin{figure}
    \centering
    \includegraphics[width=\linewidth,clip]{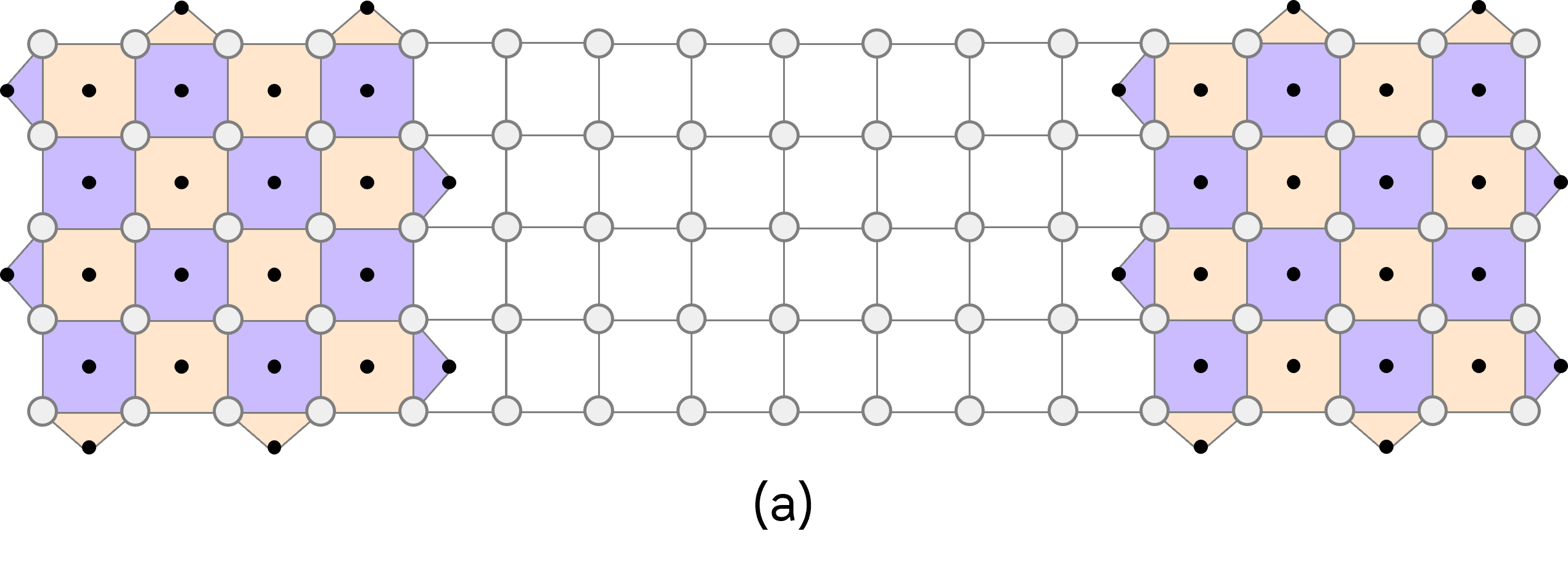}
    \includegraphics[width=\linewidth,clip]{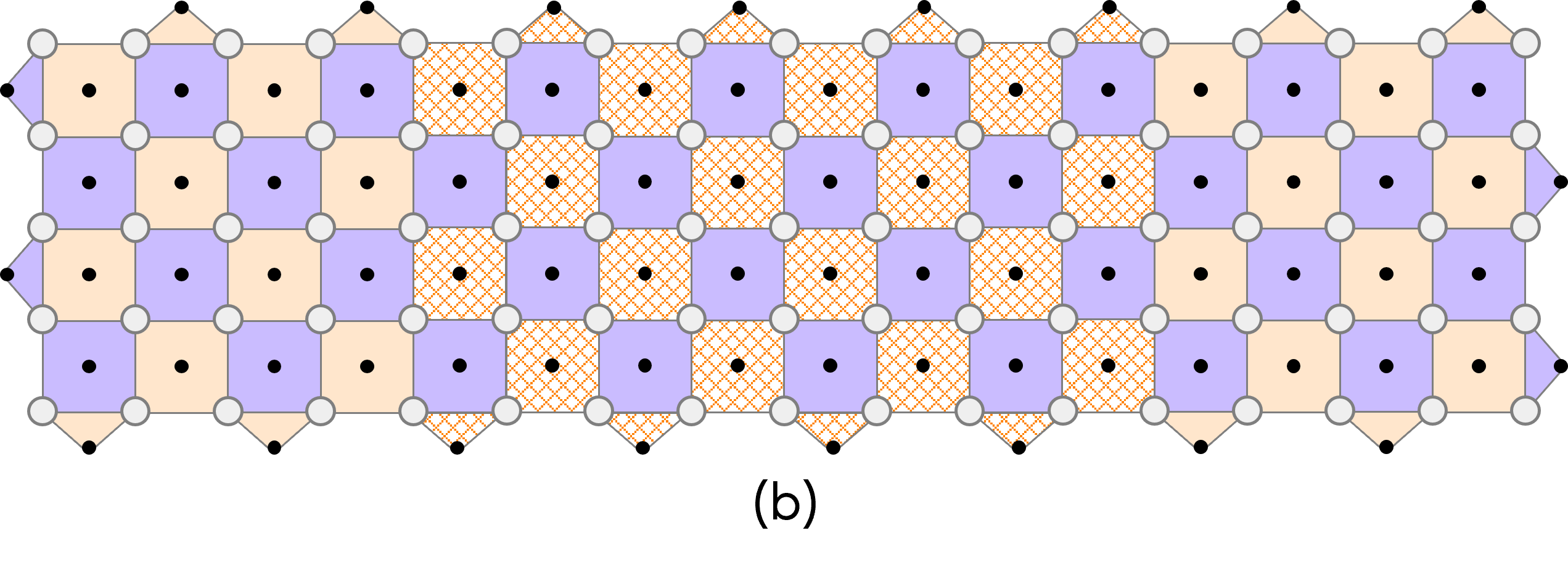}
    \caption{Surface code configuration of the $XX$ measurement experiment ($d = 5$). 
    (a) Two surface code patches before the logical $XX$ measurement. Orange (blue) faces indicate the $X$ ($Z$) stabilizers. 
    Black dots are active measurement qubits. 
    (b) Merged surface code patch. A product of the measurement values of the hatched $X$ stabilizers gives the logical $XX$ measurement result. }
    \label{fig:Hshape_conf}
\end{figure}
The qubit configuration we consider in the $XX$ measurement experiment is shown in Fig.~\ref{fig:Hshape_conf}. 
We assume the nearest neighbor connectivity between physical qubits. 
In between two surface code patch with the code distance $d$, we assume a routing space, whose width is $d + 2$. The factor 2 comes from physical qubits arranged between each surface code patch for the merging operation. 

The measurement result of the logical $XX$ operator is a product of the measurement results of the newly introduced $X$ stabilizers during the merging operation (shown as orange hatched faces in Fig.~\ref{fig:Hshape_conf}). Thus, if there are some time-like error chains in those stabilizers, the logical measurement value may flips, resulting the logical measurement error. 


\subsection{Qubit configuration of the stability experiment}
\begin{figure}
    \centering
    \includegraphics[width=0.5\linewidth,clip]{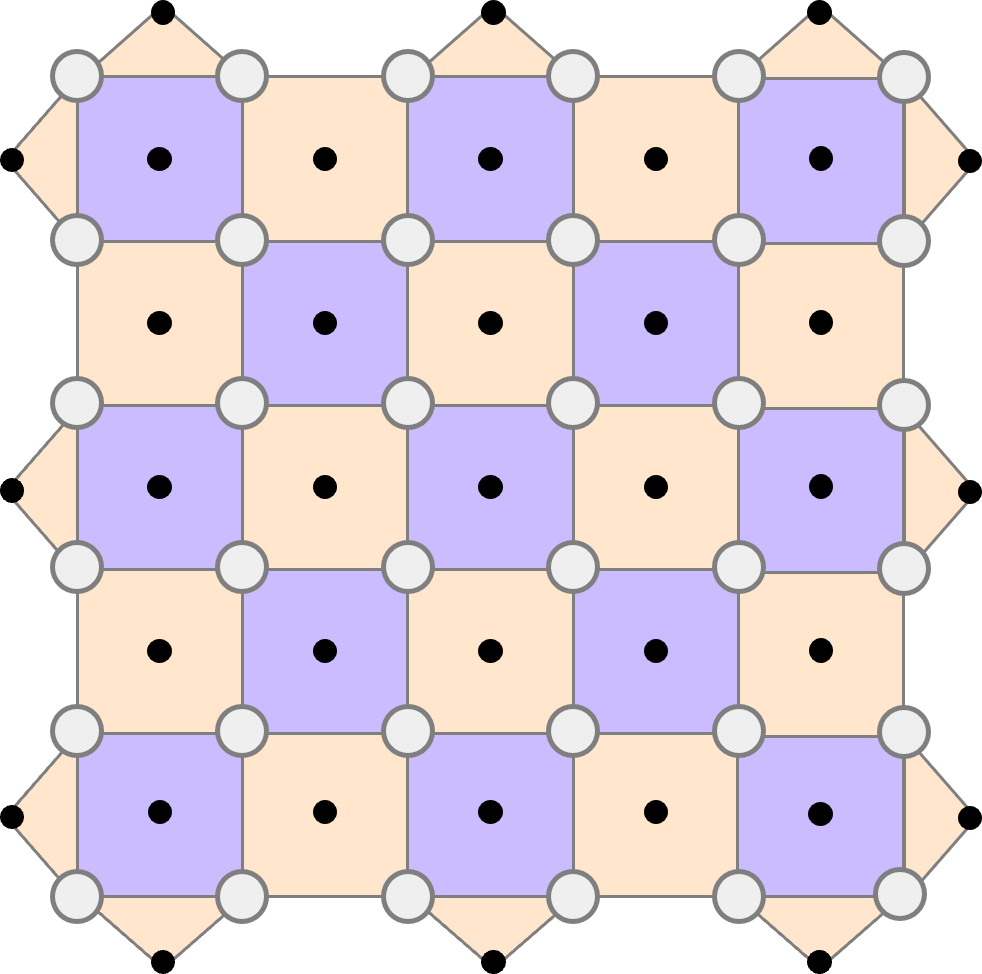}
    \caption{Surface code configuration of the stability experiment ($d = 6$). 
    A product of all measurement results of the $X$ stabilizers should be identity in this configuration if there are no error. }
    \label{fig:stability_conf}
\end{figure} 
The stability experiment~\cite{Gidney2022stability} originally introduced as the minimal experimental setup that holds the same time-like boundary condition as the routing space during the lattice surgery operation. 
In our stability experiment, we consider the $d \times d$ lattice qubit configuration with even $d$, as shown in Fig.~\ref{fig:stability_conf}. 
In this configuration, a product of all $X$ stabilizers is the identity operator; thus, a product of their measured values should be identity if there is no error. 
Time-like logical errors violate this restriction, so we can estimate the time-like logical error rate by counting such violating events from $N_{\rm sample}$ runs of this experiment. 

\section{Logical error rate fitting} \label{appx:lerfit}
\begin{figure}
    \centering
    \includegraphics[width=\linewidth,clip]{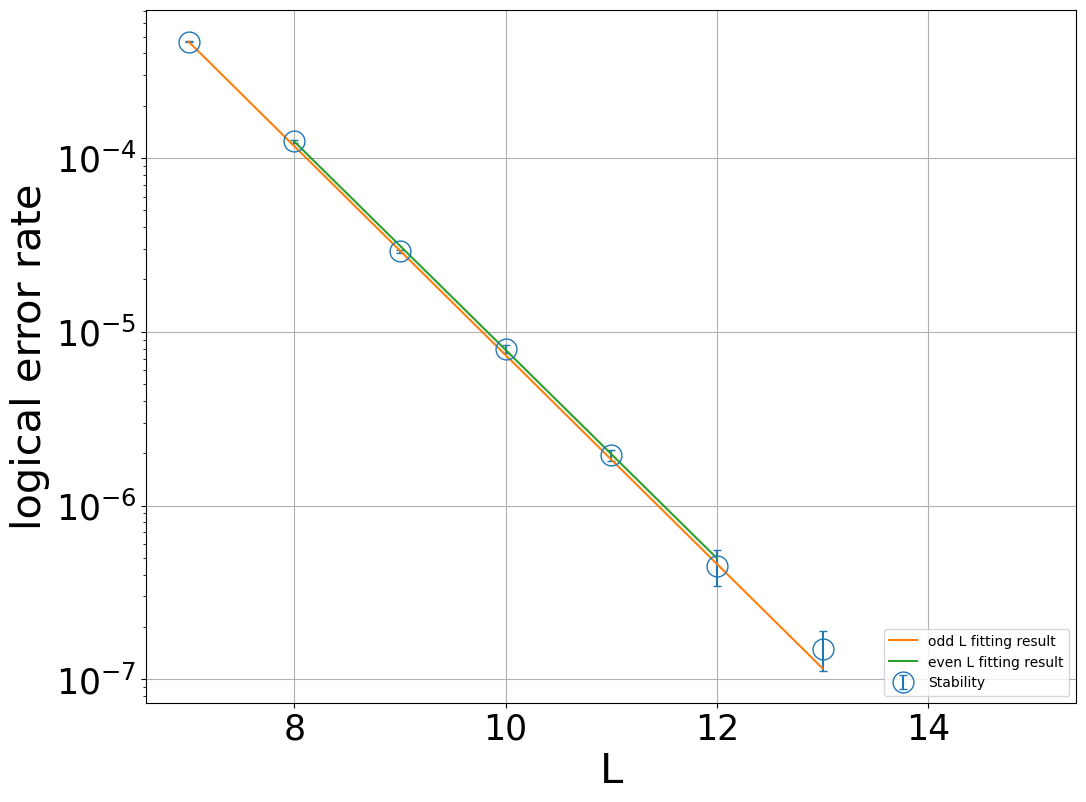}
    \caption{
    Fitting result of the logical error rate in the $d=26$ stability experiment. 
    Orange (green) line show the fitting result using odd (even) $L$ data, namely $L = 7,9,11,13$ ($L = 8,10,12$). 
    }
    \label{fig:logerr_fitting}
\end{figure}

In this appendix, we discuss the details of the fitting of the logical error rate, whose result is used for the performance estimation in Sec.~\ref{sec:accel}. 
We use the data of the stability experiment with parameters of $d = 26$, $p = 10^{-3}$. 
We fit the $L$ dependence of the logical error rate by using the fitting function, 
\begin{equation} \label{eq:logerr_fit_func}
    f(L, p) = c_1 (c_2p)^{\lfloor \frac{L+1}{2} \rfloor}, 
\end{equation}
where $c_1, c_2$ are the fitting parameters we optimize and $p = 10^{-3}$ (since we fix the spatial size of the stability experiment, we do not explicitly include the $d$ dependence in this function). 
We perform the fitting for each even and odd $L$ subset of our data.  
These results are summarized in Fig.~\ref{fig:logerr_fitting} and Tab.~\ref{tab:fit_result}. 
As seen in Tab.~\ref{tab:fit_result}, we obtain different prefactors $c_1$ for these two results, while $c_2$ takes almost the same value for each case. 
We expect that it mainly comes from the difference of the combinatorial number of the error configurations that contribute the leading logical error rate. 
We use these fitting results for estimating the target logical error rate in our performance analyses. 
\begin{table}[h]
    \centering
    \begin{tabular}{c|cc}
             & $c_1$ & $c_2$ \\ \hline
       even  & 7.80226959  & 63.20320489 \\
       odd   & 30.14749962 & 62.75117043 
    \end{tabular}
    \caption{The optimized parameters of the logical error rate fitting.}
    \label{tab:fit_result}
\end{table}

\section{Determination of $g_{\rm th}$ and $p_{\rm fail}$ in Sec.~\ref{sec:simpleex}} \label{appx:g_thdet}
\begin{figure*}
    \includegraphics[width=0.45\linewidth,clip]{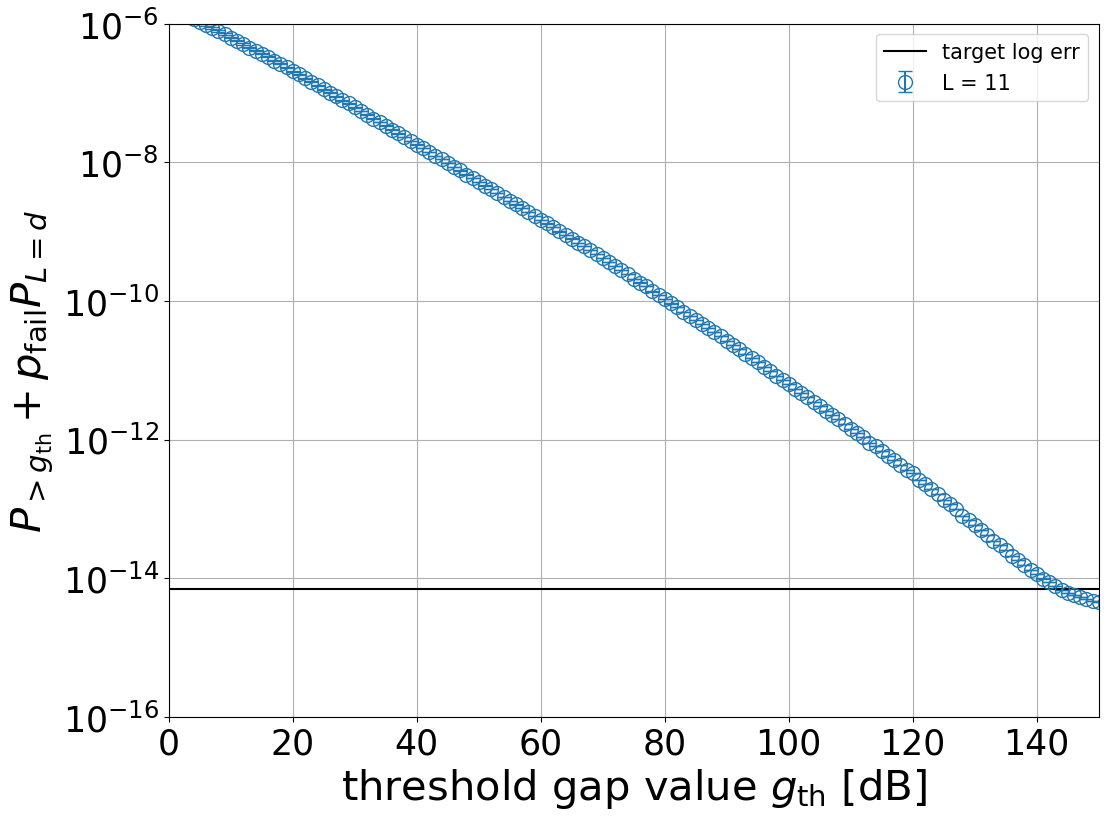} 
    \includegraphics[width=0.45\linewidth,clip]{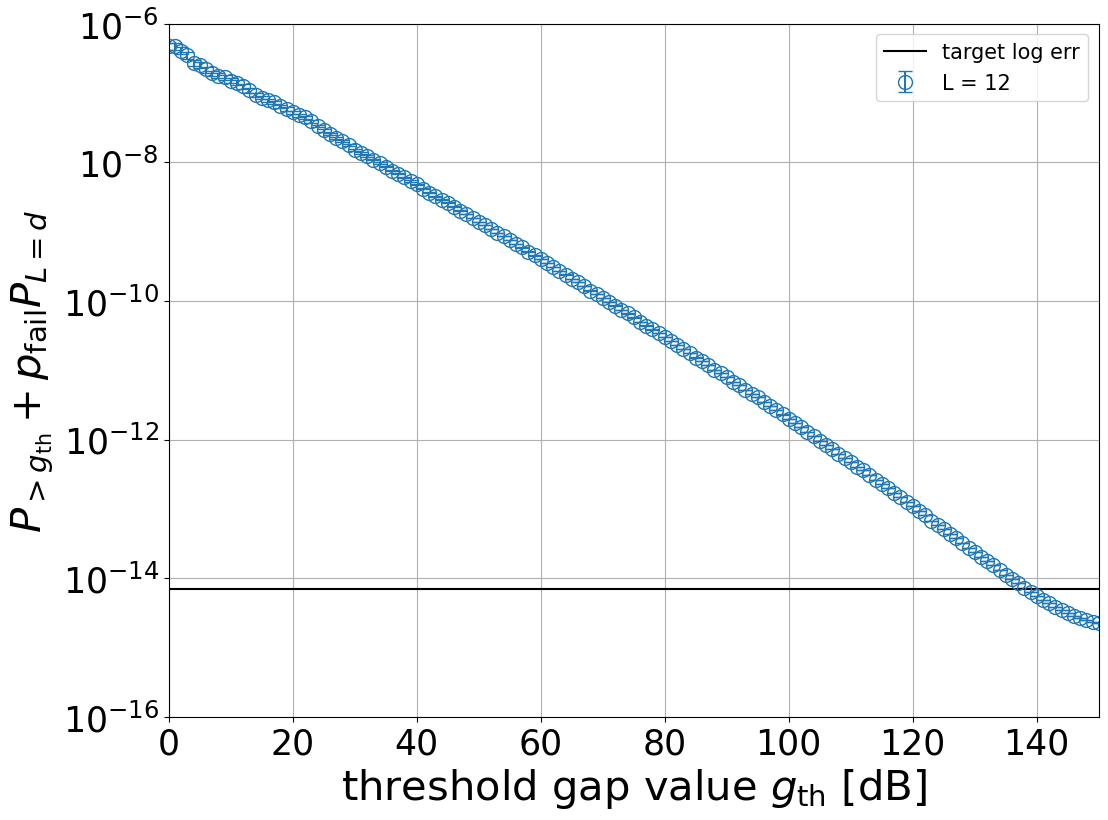} \\
    \includegraphics[width=0.45\linewidth,clip]{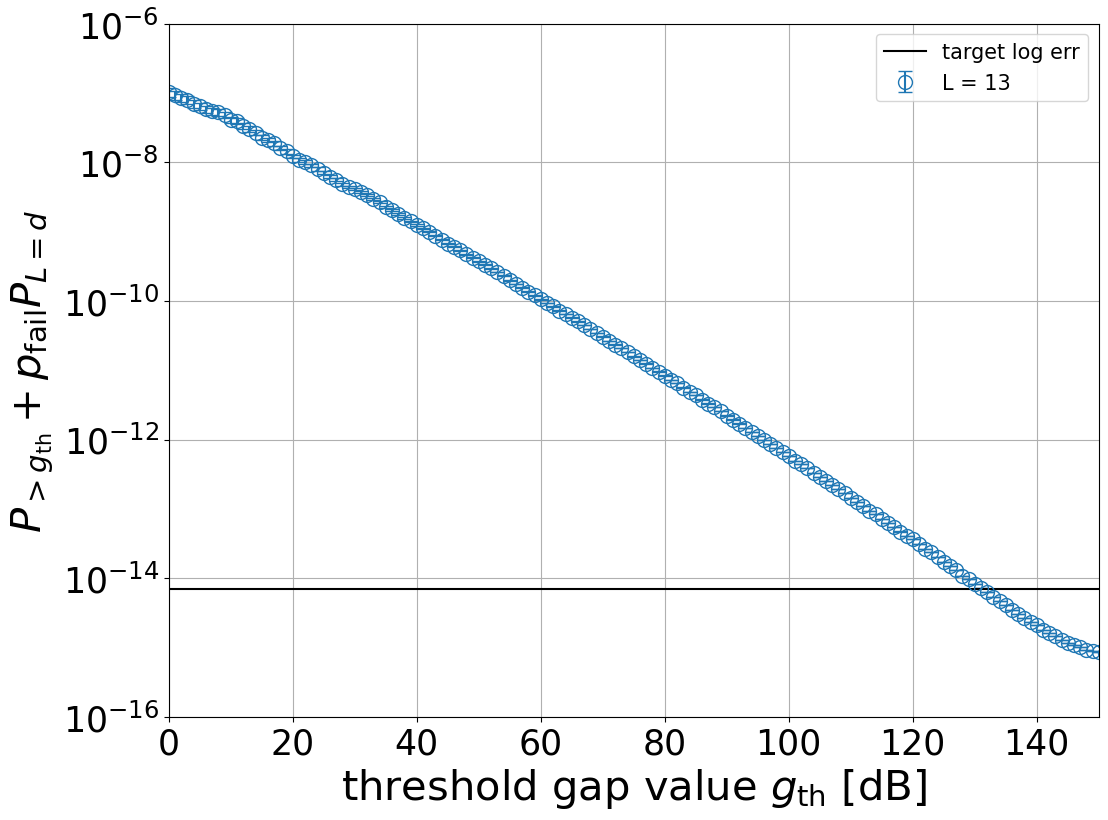} 
    \includegraphics[width=0.45\linewidth,clip]{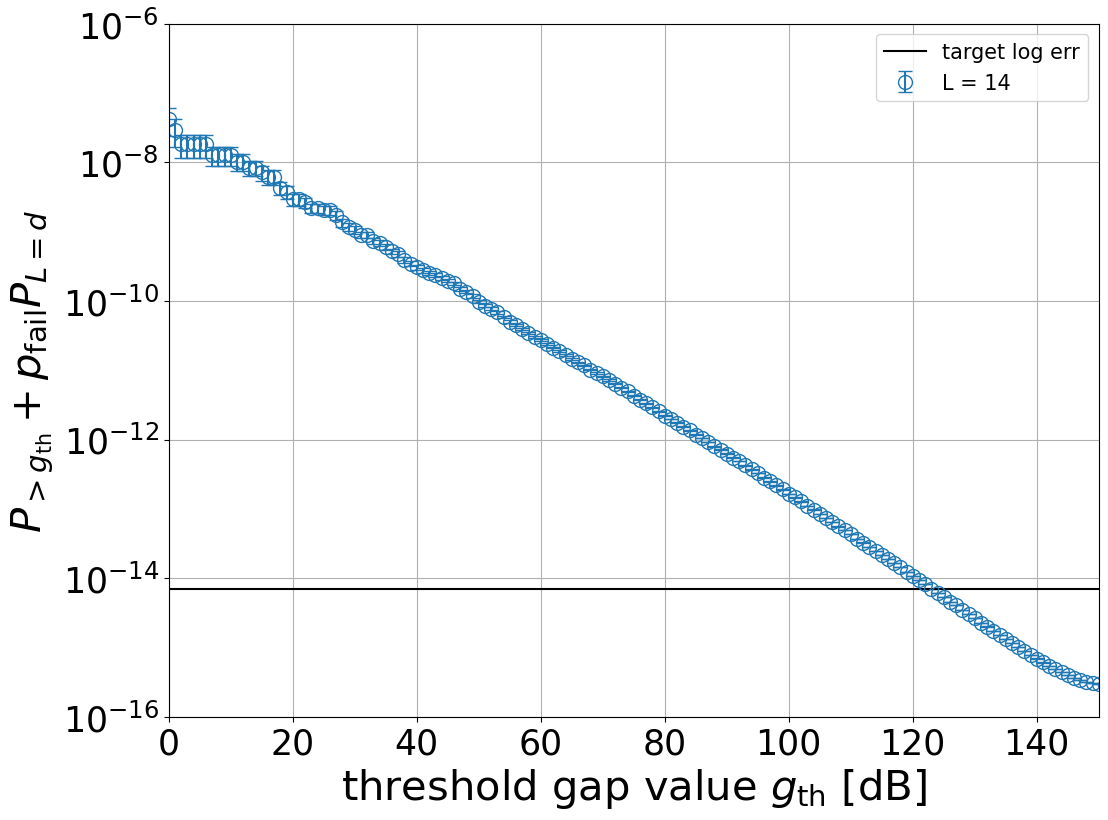} \\
    \includegraphics[width=0.45\linewidth,clip]{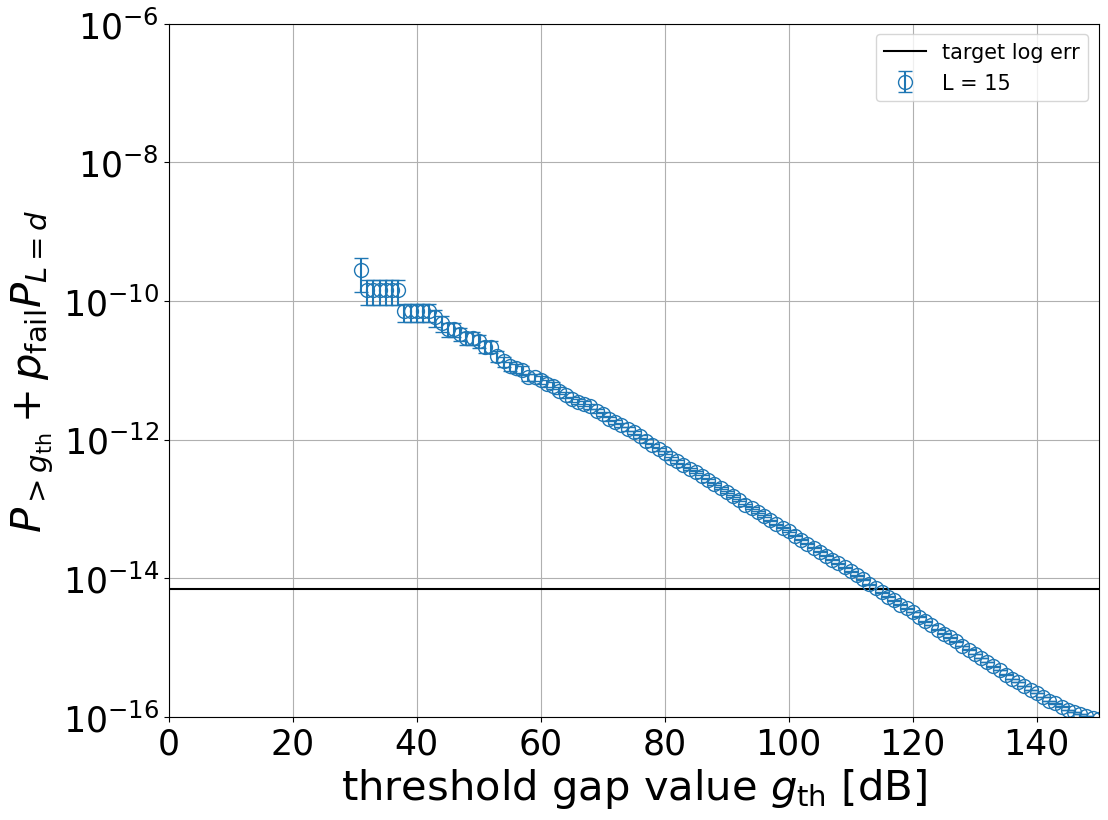} 
    \caption{
    Behavior of the total logical error rate of our protocol against $g_{\rm th}$ for $L = {11,12,13,14,15}$. 
    We show the target logical error rate with $L = d =25$ as black solid line, therefore the crossing point between blue plots and the black solid line indicates the optimal $g_{\rm th}$. 
    }
    \label{fig:determine_gth}
\end{figure*}

Here we show how to determine optimal $g_{\rm th}$ and corresponding $p_{\rm fail}$. 
As discussed, the optimal $g_{\rm th}$ is determined by solving the following condition,
\begin{eqnarray} \label{eq:gth_cond}
    P_{L=d} &=& P_{>g_{\rm th}} + p_{\rm fail} P_{L=d} \\
    &=& \int_{g_{\rm th}}^{\infty} dg P({\mathcal T}|g) P(g) + P_{L=d} \int_{0}^{g_{\rm th}} dg P(g), \nonumber
\end{eqnarray}
where $P_{L=d}$ is the target logical error rate with $L=d$ syndrome measurement rounds, which can be estimated by the fitting result discussed in Appendix~\ref{appx:lerfit}. 
The thresholded logical error rate $P_{>g_{\rm th}}$ and $p_{\rm fail}$ can be estimated by the fitting result of $P({\mathcal T}|g)$ (Fig.~\ref{fig:cond_prob}) and the gap distribution (Fig.~\ref{fig:gap_dist}). 
We plot the $g_{\rm th}$ dependence of the right hand side of Eq.~(\ref{eq:gth_cond}) for different $L$ in Fig.~\ref{fig:determine_gth}. 
It decreases when $g_{\rm th}$ increases as expected, 
and we can find an optimal $g_{\rm th}$ value satisfying the condition Eq.~(\ref{eq:gth_cond}) (crossing point between blue plots and the black solid line in Fig.~\ref{fig:determine_gth}). 
Values of $p_{\rm fail}$ at the optimal $g_{\rm th}$ are summarized in Fig.~\ref{fig:pfail_simpleXX}. 

\section{Details of the performance comparison} \label{appx:comp_details}
In this appendix, we discuss the details of the performance comparison between our proposal, TELS, and STELS. 
Firstly, we introduce the logical patch arrangement we consider in the comparison, then discuss the details of the performance estimation for each protocol. 

\subsection{Logical patch configuration}
\begin{figure}
    \centering
    \includegraphics[width=0.9\linewidth]{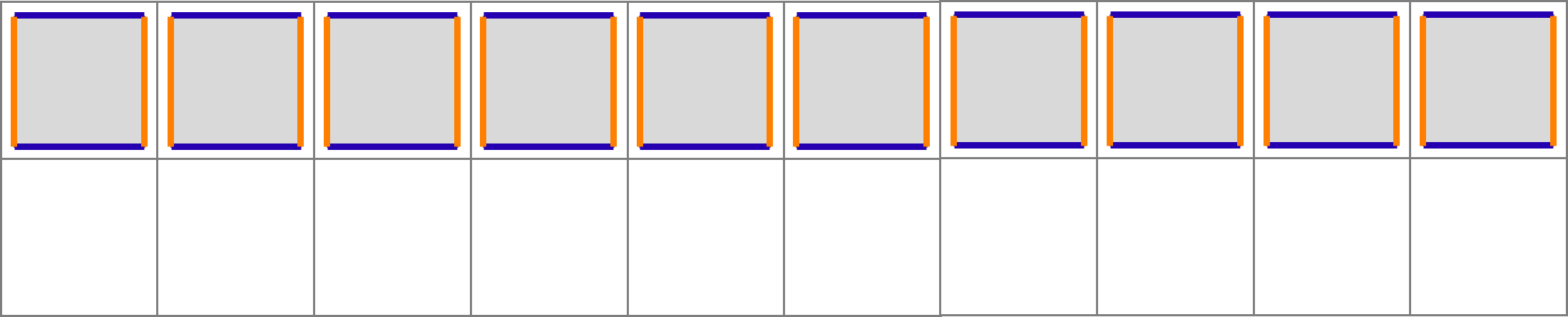}
    \caption{Logical patch configuration we consider in this study. We use a linear arrangement using $2 \times N$ patches. We place data logical patches on the first row (gray patches), and patches in the second row are utilized as the routing region (white squares).}
    \label{fig:patch_configuration_linear}
\end{figure}
We assume that the logical patches are arranged linearly ($2 \times N$), as shown in Fig.~\ref{fig:patch_configuration_linear}. 
In this configuration, we can measure the logical $Z$ operators for each patch directly via the ancilla region. 
We do not consider any overhead of basis changes (such as the patch rotation operation etc.) in the performance comparison, since this does not contribute the difference of the performance; thus, we assume to perform a chain of multi-$Z$ Pauli measurements, for example. 
We do not think that this setup is unreal. A typical example is the magic state distillation~\cite{Litinski2019gameofsurfacecodes}. 

The code distance we consider is $d = 25$ and the physical error rate is assumed to be $p = 10^{-3}$. We take $N = 10$. 
We estimate the average runtime per logical Pauli measurement for $k < 20$ by using the numerical results of the stability experiment with $d = 26$. 
Below, we describe the details of the estimation for each protocol. 

\subsection{Performance estimation for the time-like gap protocol} \label{appx:comp_ours_detail}
\begin{figure}
    \centering
    \includegraphics[width=0.8\linewidth]{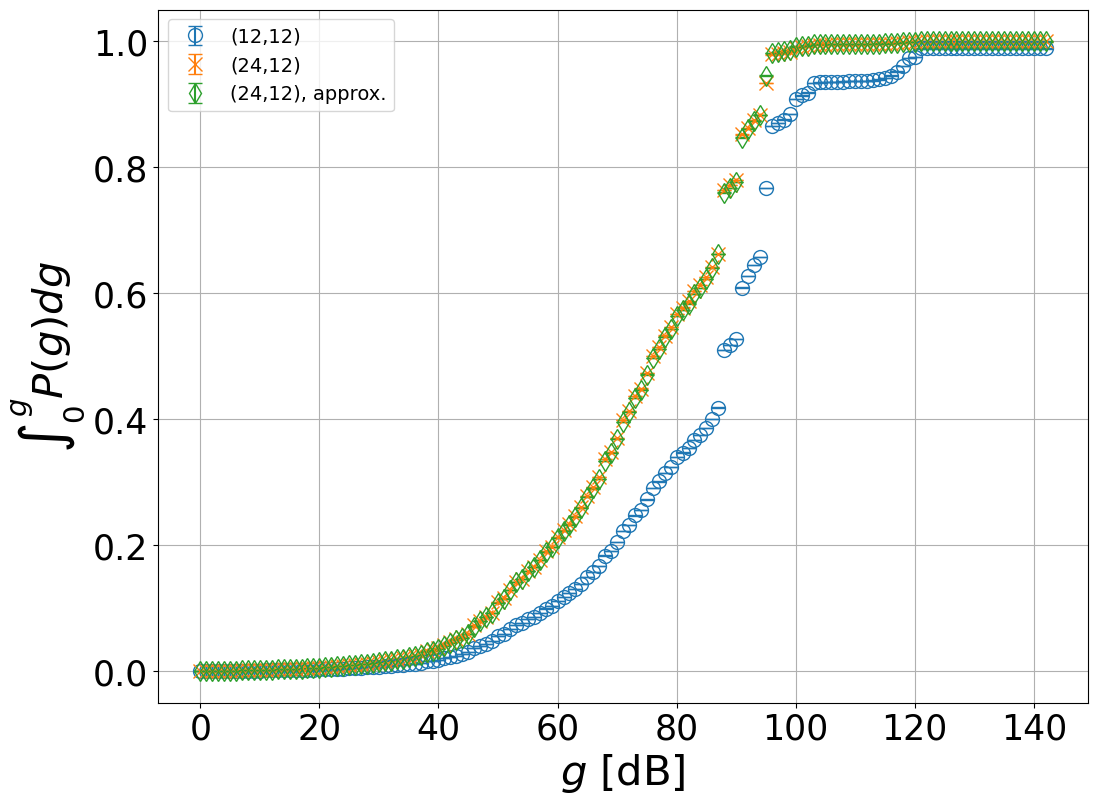}
    \caption{Cumulative probability of time-like gap distribution for the $d=12, L = 6$ stability experiment. Blue circle (orange cross) points show the $d\times d$ ($2d \times d$) grid size result. We also show the approximated $2d \times d$ result as green diamond points, which are estimated by $1 - \left(1 - \int_0^g dg P_{d\times d}(g) \right)^2$.}
    \label{fig:accum_gap_dist}
\end{figure}

As discussed in Sec.~\ref{sec:simpleex}, we firstly estimate $p_{\rm fail}$ for several $L$. 
$p_{\rm fail}$ for the single patch area can be directly estimated by the numerical results, but in the lattice surgery, the routing space includes several ancilla patches, and therefore the chance of the time-like logical error becomes larger. 
To estimate the area dependence of the time-like gap distribution, we numerically simulate two stability experiments with a grid size $d \times d$ and $2d \times d$. 
We show the example of the cumulative probability of the time-like gap distribution, $\int_0^g dg P(g)$, in Fig.\ref{fig:accum_gap_dist}. 
As seen in Fig.~\ref{fig:accum_gap_dist}, we observe that the approximated $2d \times d$ distribution, estimated by $1 - \left(1 - \int_0^g dg P_{d\times d}(g) \right)^2$, where $P_{d\times d}(g)$ is the gap distribution obtained from the $d\times d$ simulation, 
has a good agreement with the directly calculated $2d \times d$ result. Note that this kind of agreement is also reported in the context of the space-like gap distribution~\cite{gidney2023yoked}. 
Based on this fact, we assume that we can approximate the cumulative probability of the gap distribution spreading over $n$ ancilla patches by 
\begin{equation}
    \int_0^g dg P_{nd \times d}(g) = 1 - \left(1 - \int_0^g dg P_{d\times d}(g) \right)^n, 
\end{equation}
and therefore, 
\begin{equation} \label{eq:approx_gap_dist}
    P_{nd \times d}(g) = n P_{d\times d}(g) \left(1 - \int_0^g dg P_{d\times d}(g) \right)^{n-1}. 
\end{equation}
Since the conditional probability $P({\mathcal T}|g)$ is independent on the grid size, by using Eqs.(\ref{eq:cont_prob_scaling}) and (\ref{eq:approx_gap_dist}), we can determine $g_{\rm th}$ and $p_{\rm fail}$ in a same manner as discussed in Appendix~\ref{appx:g_thdet}. 
Although the area of the routing region (number of ancilla patches, $n$) depends on the weight of the Pauli operator we measure, in our estimation, we take the worst case, namely $n = N$, for all Pauli measurements. 

For the average runtime estimation, we assume that we can decode syndromes from a logical Pauli measurement during performing the next Pauli measurement; thus, the reaction time of the sequence of Pauli measurements only appears after the last measurement. 

\subsection{Performance estimation for the TELS} \label{appx:comp_tels_detail}
\begin{figure*}
    \includegraphics[width=0.45\linewidth,clip]{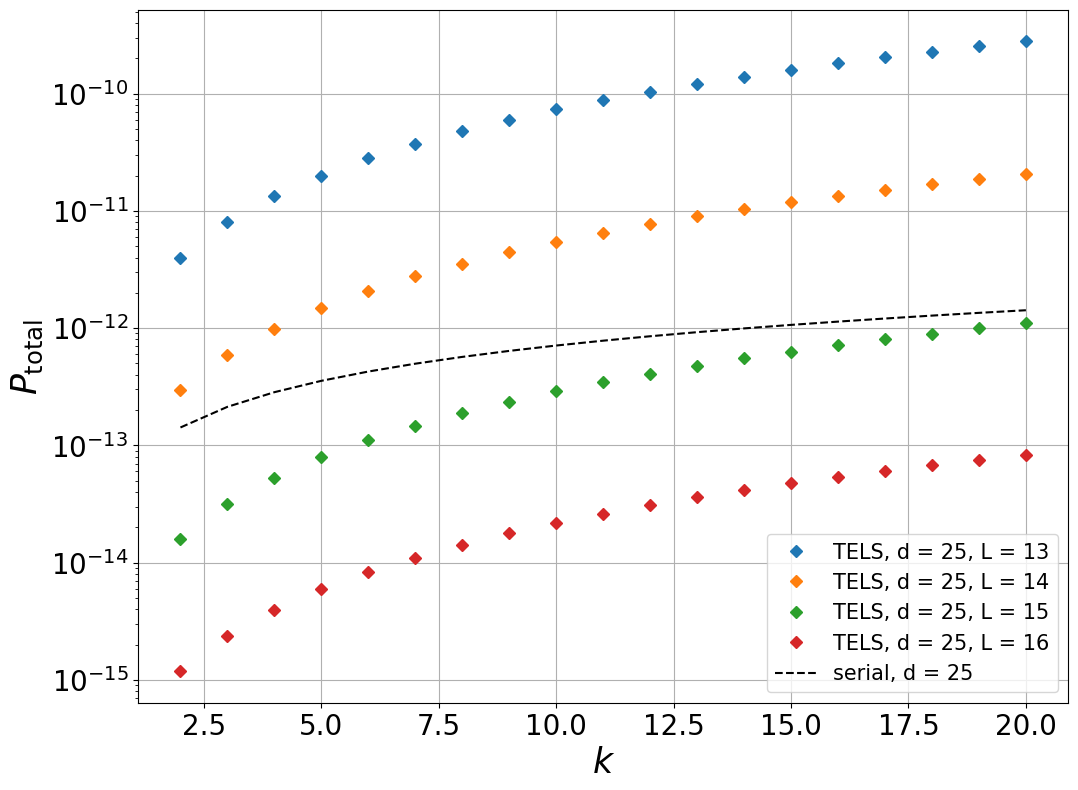} 
    \includegraphics[width=0.45\linewidth,clip]{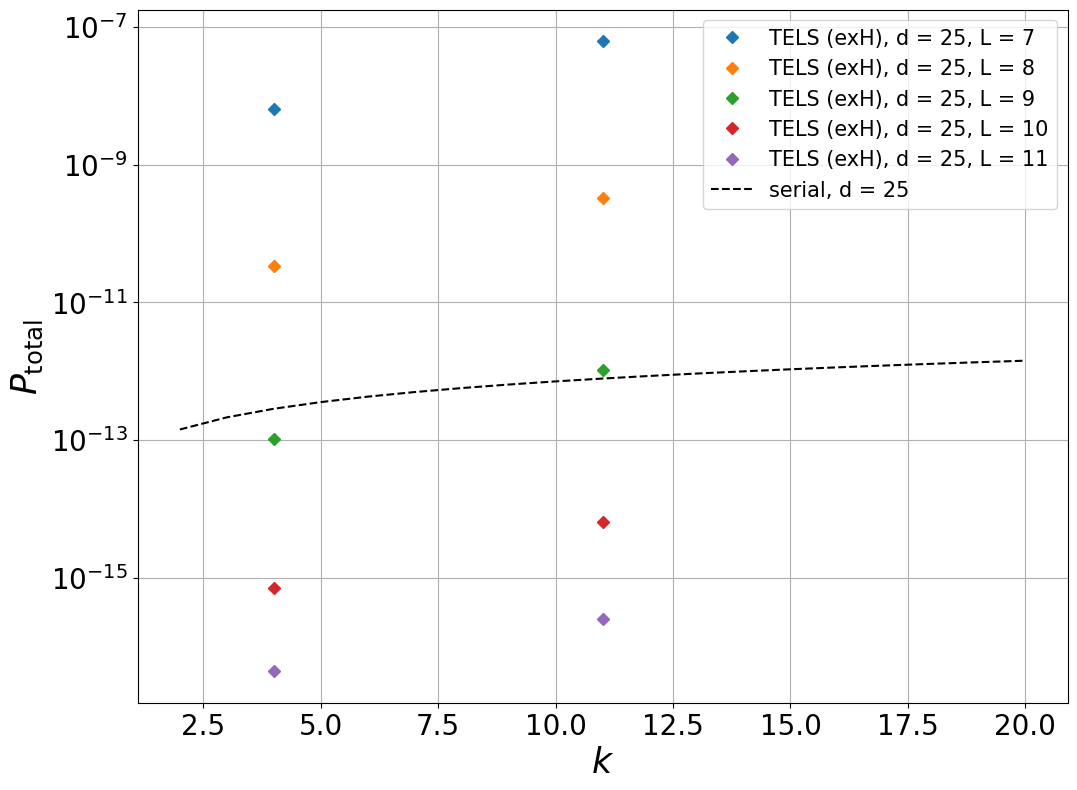} \\
    \includegraphics[width=0.45\linewidth,clip]{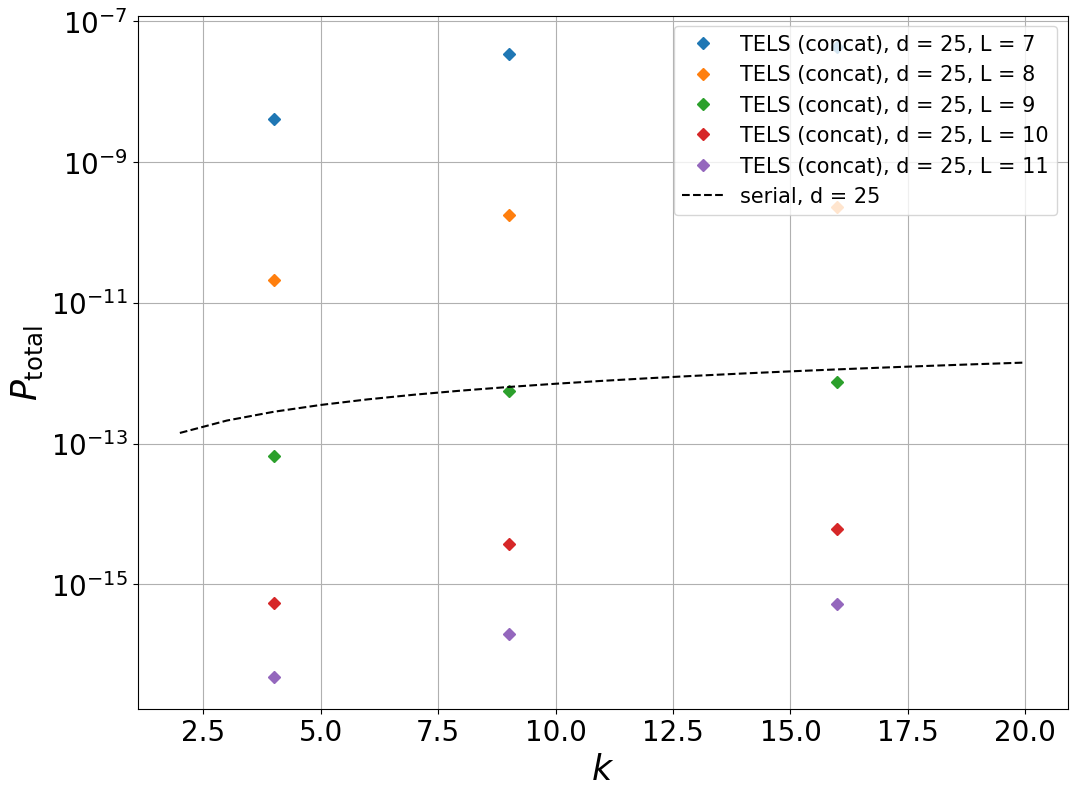} 
    \caption{
    $(L, k)$ dependence of the total logical error rate in TELS protocol. 
    We show the target total logical error rate as the dashed black lines. 
    (Top left) $[k+1, k, 2]$ measurement code. $L = 15$ is optimal. 
    (Top right) $[2^r, 2^r - r -1, 4]$ measurement code. $(L,k) = (9,4), (10, 11)$ are optimal. 
    (Bottom) $[(r+1)^2, r^2, 4]$ measurement code. $L = 9$ is optimal for all possible $k$. 
    }
    \label{fig:tels_optL}
\end{figure*}

To estimate the average runtime for the TELS protocol, we need (i) failure rate of the error detection mode, and (ii) logical error rate of the entire protocol. 
The failure rate of the error detection using a $[n,k,d]$ measurement code is given as 
\begin{equation}
    p_{\rm fail, TELS} = \sum_i C_i P^i_{L} (1-P_L)^{n-i}, 
\end{equation}
where $C_i \ (i = 1,2,...,n)$ is the number of detectable error patterns consisting of $i$ errors, and $P_L$ is the logical error rate of a single Pauli measurement with $L$ syndrome cycles. 
In our analysis, we estimate $P_L$ by using the fitting result discussed in Appendix~\ref{appx:lerfit}. 
The fitting result corresponds to the case where the routing region is a single patch, 
but since the leading-order contribution of the logical error rate is proportional to the area of the routing region, 
we can estimate the logical error rate for routing region spreading over $n$ ancilla patches as $P_L = 1 - (1-P_{L, {\rm single}})^n \approx n P_{L, {\rm single}}$, where $P_{L,{\rm single}}$ is the logical error rate for the single patch area. 
Same as the discussion in Appendix~\ref{appx:comp_ours_detail}, we consider the worst case failure rate. 
Therefore, for the TELS using a $[n, k, d]$ measurement code, 
\begin{equation} 
    p_{\rm fail, TELS} = \sum_i C_i (NP_{L,{\rm single}})^i (1-NP_{L, {\rm single}})^{n-i}. 
\end{equation}

Similarly, we can estimate the logical error rate of the TELS protocol. 
The general $[n, k, d]$ code in the error detection mode cannot detect some error patterns consisting of $d$ errors. 
By considering the logical error rate of the remeasurement, we obtain the entire logical error rate as 
\begin{equation}
    P_{\rm TELS} = p_{\rm fail, TELS} (1-(1-P_{d})^k) + \bar C_d P_L^d (1-P_L)^{n-d}, 
\end{equation}
where $P_{d}$ is the logical error rate of a Pauli measurement with $d$ syndrome cycles, and $\bar C_d$ is the number of undetectable error patterns. Clearly, the first term comes from the remeasurement branch and the second term is the contribution from the undetectable error patterns. $P_d$ and $P_L$ in this equation are estimated by the worst-case assumption, as discussed above. 

Once we fix the measurement code we utilize, the optimal choice of $L$ is determined by the condition 
\begin{equation}
    P_{\rm TELS} \leq 1-(1-P_{d})^k. 
\end{equation}
We summarize the $P_{\rm TELS}$ for several $L$ and $k$ in Fig.~\ref{fig:tels_optL}. 

For the average runtime estimation, we assume that the reaction time of the Pauli measurements only appears after the last measurement, as discussed in ~\ref{appx:comp_ours_detail}. 

\subsection{Performance estimation for the STELS} \label{appx:stels_gth_determination}
The performance estimation for the STELS protocol is straightforward, just combining the discussion in Appendix~\ref{appx:comp_ours_detail} and \ref{appx:comp_tels_detail}. 
First of all, the optimization of $L_{\rm TELS}$ is done by the same procedure discussed in Appendix~\ref{appx:comp_tels_detail}. 
Once $L_{\rm TELS}$ is fixed, we next optimize the time-like gap method parameter, $L' < L_{\rm TELS}$. 
This optimization is the same procedure discussed in Appendix~\ref{appx:comp_ours_detail}, except that the target logical error rate is estimated by using $L_{\rm TELS}$ measurement cycles, not $d$ measurement cycles. 
The condition to determine optimal $g_{\rm th}$ is 
\begin{eqnarray} \label{eq:gth_det_stels} 
    P_{L_{\rm TELS}} &=& P_{>g_{\rm th}} + p_{\rm fail} P_{L_{\rm TELS}}, 
\end{eqnarray}
where $P_{L_{\rm TELS}}$ is the time-like logical error rate with $L = L_{\rm TELS}$. 
We search several $L'$ to find smallest average runtime. 

\section{Runtime reduction in the logical CNOT gate} \label{appx:cnot_circuit}
\begin{figure}
    \centering
    \mbox{
        \Qcircuit @C=1em @R=.7em {
        \lstick{\ket{c}} & \multigate{1}{M_{ZZ}(m_1)} & \qw & \gate{Z^{m_2}} & \qw \\
        \lstick{\ket{+}} & \ghost{M_{ZZ}(m_1)} & \multigate{1}{M_{XX}(m_2)} & \gate{M_Z(m_3)} \\
        \lstick{\ket{t}} & \qw  & \ghost{M_{XX}(m_2)} & \gate{X^{m_1 + m_3}} & \qw
        }
    }    
    \caption{Logical CNOT gate in the lattice surgery. $\ket{c}$ and $\ket{t}$ indicate control and target qubits in the logical CNOT operation.
    $m_i$ ($i = 1,2,3$) are measured parities.}
    \label{fig:cnot_circuit}
\end{figure}
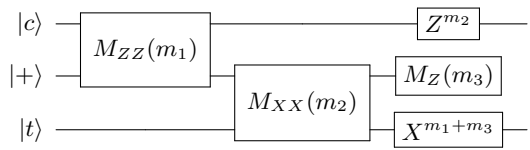

In this appendix, we shortly explain the application of our protocol to the logical CNOT gate via the lattice surgery. 
The logical CNOT gate is typically performed by the sequence of the $ZZ$, $XX$, and $Z$ measurements via a single ancilla patch, as shown in Fig.~\ref{fig:cnot_circuit}. 
By considering the naive lattice surgery protocol, each logical two-Pauli measurement takes $d$ rounds of the syndrome measurements. In this sequence, we do not need to wait for the logical measurement result of the $ZZ$ measurement before starting the $XX$ measurement, since the Pauli correction depending on the measurement result can be tracked by the Pauli frame later. 
Therefore, the total runtime of this sequence is just $2d$ (we omit the constant contribution from the destructive $Z$ measurement). 

Next, let us consider to apply our protocol in this circuit. 
We can apply it to both $ZZ$ and $XX$ measurements. 
In this case, however, to fix the time-like logical gap in the first $ZZ$ measurement, we have to wait before moving on to the next $XX$ measurement, since $ZZ$ and $XX$ measurement do not commute with each other. 
The same situation occurs in the next $XX$ measurement. 
Therefore, we take at least $2(L+T)$ for finishing this circuit (we assume $p_{\rm fail} \approx 0$ here). Comparing to the naive case, our protocol is advantageous when $L+T < d$ is satisfied. 
Fortunately, if we use the assumption used in the main text, namely $d = 25, p = 10^{-3}, T = 10$, the optimal $L$ is 13, so $L+T = 23 < d = 25$ is satisfied, but its reduction ratio is just 8\%. 

\bibliography{bib.bib}

\begin{thebibliography}{49}%
\makeatletter
\providecommand \@ifxundefined [1]{%
 \@ifx{#1\undefined}
}%
\providecommand \@ifnum [1]{%
 \ifnum #1\expandafter \@firstoftwo
 \else \expandafter \@secondoftwo
 \fi
}%
\providecommand \@ifx [1]{%
 \ifx #1\expandafter \@firstoftwo
 \else \expandafter \@secondoftwo
 \fi
}%
\providecommand \natexlab [1]{#1}%
\providecommand \enquote  [1]{``#1''}%
\providecommand \bibnamefont  [1]{#1}%
\providecommand \bibfnamefont [1]{#1}%
\providecommand \citenamefont [1]{#1}%
\providecommand \href@noop [0]{\@secondoftwo}%
\providecommand \href [0]{\begingroup \@sanitize@url \@href}%
\providecommand \@href[1]{\@@startlink{#1}\@@href}%
\providecommand \@@href[1]{\endgroup#1\@@endlink}%
\providecommand \@sanitize@url [0]{\catcode `\\12\catcode `\$12\catcode `\&12\catcode `\#12\catcode `\^12\catcode `\_12\catcode `\%12\relax}%
\providecommand \@@startlink[1]{}%
\providecommand \@@endlink[0]{}%
\providecommand \url  [0]{\begingroup\@sanitize@url \@url }%
\providecommand \@url [1]{\endgroup\@href {#1}{\urlprefix }}%
\providecommand \urlprefix  [0]{URL }%
\providecommand \Eprint [0]{\href }%
\providecommand \doibase [0]{https://doi.org/}%
\providecommand \selectlanguage [0]{\@gobble}%
\providecommand \bibinfo  [0]{\@secondoftwo}%
\providecommand \bibfield  [0]{\@secondoftwo}%
\providecommand \translation [1]{[#1]}%
\providecommand \BibitemOpen [0]{}%
\providecommand \bibitemStop [0]{}%
\providecommand \bibitemNoStop [0]{.\EOS\space}%
\providecommand \EOS [0]{\spacefactor3000\relax}%
\providecommand \BibitemShut  [1]{\csname bibitem#1\endcsname}%
\let\auto@bib@innerbib\@empty
\bibitem [{\citenamefont {Shor}(1999)}]{shor1999polynomial}%
  \BibitemOpen
  \bibfield  {author} {\bibinfo {author} {\bibfnamefont {P.~W.}\ \bibnamefont {Shor}},\ }\bibfield  {title} {\bibinfo {title} {Polynomial-time algorithms for prime factorization and discrete logarithms on a quantum computer},\ }\href {https://doi.org/10.1137/S0097539795293172} {\bibfield  {journal} {\bibinfo  {journal} {SIAM review}\ }\textbf {\bibinfo {volume} {41}},\ \bibinfo {pages} {303} (\bibinfo {year} {1999})}\BibitemShut {NoStop}%
\bibitem [{\citenamefont {Abrams}\ and\ \citenamefont {Lloyd}(1999)}]{abrams1999quantum}%
  \BibitemOpen
  \bibfield  {author} {\bibinfo {author} {\bibfnamefont {D.~S.}\ \bibnamefont {Abrams}}\ and\ \bibinfo {author} {\bibfnamefont {S.}~\bibnamefont {Lloyd}},\ }\bibfield  {title} {\bibinfo {title} {Quantum algorithm providing exponential speed increase for finding eigenvalues and eigenvectors},\ }\href {https://doi.org/10.1103/PhysRevLett.83.5162} {\bibfield  {journal} {\bibinfo  {journal} {Physical Review Letters}\ }\textbf {\bibinfo {volume} {83}},\ \bibinfo {pages} {5162} (\bibinfo {year} {1999})}\BibitemShut {NoStop}%
\bibitem [{\citenamefont {Aspuru-Guzik}\ \emph {et~al.}(2005)\citenamefont {Aspuru-Guzik}, \citenamefont {Dutoi}, \citenamefont {Love},\ and\ \citenamefont {Head-Gordon}}]{aspuru2005simulated}%
  \BibitemOpen
  \bibfield  {author} {\bibinfo {author} {\bibfnamefont {A.}~\bibnamefont {Aspuru-Guzik}}, \bibinfo {author} {\bibfnamefont {A.~D.}\ \bibnamefont {Dutoi}}, \bibinfo {author} {\bibfnamefont {P.~J.}\ \bibnamefont {Love}},\ and\ \bibinfo {author} {\bibfnamefont {M.}~\bibnamefont {Head-Gordon}},\ }\bibfield  {title} {\bibinfo {title} {Simulated quantum computation of molecular energies},\ }\href {https://doi.org/10.1126/science.1113479} {\bibfield  {journal} {\bibinfo  {journal} {Science}\ }\textbf {\bibinfo {volume} {309}},\ \bibinfo {pages} {1704} (\bibinfo {year} {2005})}\BibitemShut {NoStop}%
\bibitem [{\citenamefont {Harrow}\ \emph {et~al.}(2009)\citenamefont {Harrow}, \citenamefont {Hassidim},\ and\ \citenamefont {Lloyd}}]{harrow2009quantum}%
  \BibitemOpen
  \bibfield  {author} {\bibinfo {author} {\bibfnamefont {A.~W.}\ \bibnamefont {Harrow}}, \bibinfo {author} {\bibfnamefont {A.}~\bibnamefont {Hassidim}},\ and\ \bibinfo {author} {\bibfnamefont {S.}~\bibnamefont {Lloyd}},\ }\bibfield  {title} {\bibinfo {title} {Quantum algorithm for linear systems of equations},\ }\href {https://doi.org/10.1103/PhysRevLett.103.150502} {\bibfield  {journal} {\bibinfo  {journal} {Physical review letters}\ }\textbf {\bibinfo {volume} {103}},\ \bibinfo {pages} {150502} (\bibinfo {year} {2009})}\BibitemShut {NoStop}%
\bibitem [{\citenamefont {Kitaev}(2003)}]{KITAEV20032}%
  \BibitemOpen
  \bibfield  {author} {\bibinfo {author} {\bibfnamefont {A.}~\bibnamefont {Kitaev}},\ }\bibfield  {title} {\bibinfo {title} {Fault-tolerant quantum computation by anyons},\ }\href {https://doi.org/https://doi.org/10.1016/S0003-4916(02)00018-0} {\bibfield  {journal} {\bibinfo  {journal} {Annals of Physics}\ }\textbf {\bibinfo {volume} {303}},\ \bibinfo {pages} {2} (\bibinfo {year} {2003})}\BibitemShut {NoStop}%
\bibitem [{\citenamefont {Dennis}\ \emph {et~al.}(2002)\citenamefont {Dennis}, \citenamefont {Kitaev}, \citenamefont {Landahl},\ and\ \citenamefont {Preskill}}]{10.1063/1.1499754}%
  \BibitemOpen
  \bibfield  {author} {\bibinfo {author} {\bibfnamefont {E.}~\bibnamefont {Dennis}}, \bibinfo {author} {\bibfnamefont {A.}~\bibnamefont {Kitaev}}, \bibinfo {author} {\bibfnamefont {A.}~\bibnamefont {Landahl}},\ and\ \bibinfo {author} {\bibfnamefont {J.}~\bibnamefont {Preskill}},\ }\bibfield  {title} {\bibinfo {title} {{Topological quantum memory}},\ }\href {https://doi.org/10.1063/1.1499754} {\bibfield  {journal} {\bibinfo  {journal} {Journal of Mathematical Physics}\ }\textbf {\bibinfo {volume} {43}},\ \bibinfo {pages} {4452} (\bibinfo {year} {2002})},\ \Eprint {https://arxiv.org/abs/https://pubs.aip.org/aip/jmp/article-pdf/43/9/4452/19183135/4452\_1\_online.pdf} {https://pubs.aip.org/aip/jmp/article-pdf/43/9/4452/19183135/4452\_1\_online.pdf} \BibitemShut {NoStop}%
\bibitem [{\citenamefont {Raussendorf}\ and\ \citenamefont {Harrington}(2007)}]{PhysRevLett.98.190504}%
  \BibitemOpen
  \bibfield  {author} {\bibinfo {author} {\bibfnamefont {R.}~\bibnamefont {Raussendorf}}\ and\ \bibinfo {author} {\bibfnamefont {J.}~\bibnamefont {Harrington}},\ }\bibfield  {title} {\bibinfo {title} {Fault-tolerant quantum computation with high threshold in two dimensions},\ }\href {https://doi.org/10.1103/PhysRevLett.98.190504} {\bibfield  {journal} {\bibinfo  {journal} {Phys. Rev. Lett.}\ }\textbf {\bibinfo {volume} {98}},\ \bibinfo {pages} {190504} (\bibinfo {year} {2007})}\BibitemShut {NoStop}%
\bibitem [{\citenamefont {Fowler}\ \emph {et~al.}(2012)\citenamefont {Fowler}, \citenamefont {Mariantoni}, \citenamefont {Martinis},\ and\ \citenamefont {Cleland}}]{PhysRevA.86.032324}%
  \BibitemOpen
  \bibfield  {author} {\bibinfo {author} {\bibfnamefont {A.~G.}\ \bibnamefont {Fowler}}, \bibinfo {author} {\bibfnamefont {M.}~\bibnamefont {Mariantoni}}, \bibinfo {author} {\bibfnamefont {J.~M.}\ \bibnamefont {Martinis}},\ and\ \bibinfo {author} {\bibfnamefont {A.~N.}\ \bibnamefont {Cleland}},\ }\bibfield  {title} {\bibinfo {title} {Surface codes: Towards practical large-scale quantum computation},\ }\href {https://doi.org/10.1103/PhysRevA.86.032324} {\bibfield  {journal} {\bibinfo  {journal} {Phys. Rev. A}\ }\textbf {\bibinfo {volume} {86}},\ \bibinfo {pages} {032324} (\bibinfo {year} {2012})}\BibitemShut {NoStop}%
\bibitem [{\citenamefont {Horsman}\ \emph {et~al.}(2012)\citenamefont {Horsman}, \citenamefont {Fowler}, \citenamefont {Devitt},\ and\ \citenamefont {Meter}}]{Horsman_2012}%
  \BibitemOpen
  \bibfield  {author} {\bibinfo {author} {\bibfnamefont {C.}~\bibnamefont {Horsman}}, \bibinfo {author} {\bibfnamefont {A.~G.}\ \bibnamefont {Fowler}}, \bibinfo {author} {\bibfnamefont {S.}~\bibnamefont {Devitt}},\ and\ \bibinfo {author} {\bibfnamefont {R.~V.}\ \bibnamefont {Meter}},\ }\bibfield  {title} {\bibinfo {title} {Surface code quantum computing by lattice surgery},\ }\href {https://doi.org/10.1088/1367-2630/14/12/123011} {\bibfield  {journal} {\bibinfo  {journal} {New Journal of Physics}\ }\textbf {\bibinfo {volume} {14}},\ \bibinfo {pages} {123011} (\bibinfo {year} {2012})}\BibitemShut {NoStop}%
\bibitem [{\citenamefont {Acharya}\ \emph {et~al.}(2024)\citenamefont {Acharya}, \citenamefont {Abanin}, \citenamefont {Aghababaie-Beni}, \citenamefont {Aleiner}, \citenamefont {Andersen}, \citenamefont {Ansmann}, \citenamefont {Arute}, \citenamefont {Arya}, \citenamefont {Asfaw}, \citenamefont {Astrakhantsev}, \citenamefont {Atalaya}, \citenamefont {Babbush}, \citenamefont {Bacon}, \citenamefont {Ballard}, \citenamefont {Bardin}, \citenamefont {Bausch}, \citenamefont {Bengtsson}, \citenamefont {Bilmes}, \citenamefont {Blackwell}, \citenamefont {Boixo}, \citenamefont {Bortoli}, \citenamefont {Bourassa}, \citenamefont {Bovaird}, \citenamefont {Brill}, \citenamefont {Broughton}, \citenamefont {Browne}, \citenamefont {Buchea}, \citenamefont {Buckley}, \citenamefont {Buell}, \citenamefont {Burger}, \citenamefont {Burkett}, \citenamefont {Bushnell}, \citenamefont {Cabrera}, \citenamefont {Campero}, \citenamefont {Chang}, \citenamefont {Chen}, \citenamefont {Chen}, \citenamefont {Chiaro}, \citenamefont {Chik},
  \citenamefont {Chou}, \citenamefont {Claes}, \citenamefont {Cleland}, \citenamefont {Cogan}, \citenamefont {Collins}, \citenamefont {Conner}, \citenamefont {Courtney}, \citenamefont {Crook}, \citenamefont {Curtin}, \citenamefont {Das}, \citenamefont {Davies}, \citenamefont {De~Lorenzo}, \citenamefont {Debroy}, \citenamefont {Demura}, \citenamefont {Devoret}, \citenamefont {Di~Paolo}, \citenamefont {Donohoe}, \citenamefont {Drozdov}, \citenamefont {Dunsworth}, \citenamefont {Earle}, \citenamefont {Edlich}, \citenamefont {Eickbusch}, \citenamefont {Elbag}, \citenamefont {Elzouka}, \citenamefont {Erickson}, \citenamefont {Faoro}, \citenamefont {Farhi}, \citenamefont {Ferreira}, \citenamefont {Burgos}, \citenamefont {Forati}, \citenamefont {Fowler}, \citenamefont {Foxen}, \citenamefont {Ganjam}, \citenamefont {Garcia}, \citenamefont {Gasca}, \citenamefont {Genois}, \citenamefont {Giang}, \citenamefont {Gidney}, \citenamefont {Gilboa}, \citenamefont {Gosula}, \citenamefont {Dau}, \citenamefont {Graumann},
  \citenamefont {Greene}, \citenamefont {Gross}, \citenamefont {Habegger}, \citenamefont {Hall}, \citenamefont {Hamilton}, \citenamefont {Hansen}, \citenamefont {Harrigan}, \citenamefont {Harrington}, \citenamefont {Heras}, \citenamefont {Heslin}, \citenamefont {Heu}, \citenamefont {Higgott}, \citenamefont {Hill}, \citenamefont {Hilton}, \citenamefont {Holland}, \citenamefont {Hong}, \citenamefont {Huang}, \citenamefont {Huff}, \citenamefont {Huggins}, \citenamefont {Ioffe}, \citenamefont {Isakov}, \citenamefont {Iveland}, \citenamefont {Jeffrey}, \citenamefont {Jiang}, \citenamefont {Jones}, \citenamefont {Jordan}, \citenamefont {Joshi}, \citenamefont {Juhas}, \citenamefont {Kafri}, \citenamefont {Kang}, \citenamefont {Karamlou}, \citenamefont {Kechedzhi}, \citenamefont {Kelly}, \citenamefont {Khaire}, \citenamefont {Khattar}, \citenamefont {Khezri}, \citenamefont {Kim}, \citenamefont {Klimov}, \citenamefont {Klots}, \citenamefont {Kobrin}, \citenamefont {Kohli}, \citenamefont {Korotkov}, \citenamefont
  {Kostritsa}, \citenamefont {Kothari}, \citenamefont {Kozlovskii}, \citenamefont {Kreikebaum}, \citenamefont {Kurilovich}, \citenamefont {Lacroix}, \citenamefont {Landhuis}, \citenamefont {Lange-Dei}, \citenamefont {Langley}, \citenamefont {Laptev}, \citenamefont {Lau}, \citenamefont {Le~Guevel}, \citenamefont {Ledford}, \citenamefont {Lee}, \citenamefont {Lee}, \citenamefont {Lensky}, \citenamefont {Leon}, \citenamefont {Lester}, \citenamefont {Li}, \citenamefont {Li}, \citenamefont {Lill}, \citenamefont {Liu}, \citenamefont {Livingston}, \citenamefont {Locharla}, \citenamefont {Lucero}, \citenamefont {Lundahl}, \citenamefont {Lunt}, \citenamefont {Madhuk}, \citenamefont {Malone}, \citenamefont {Maloney}, \citenamefont {Mandrà}, \citenamefont {Manyika}, \citenamefont {Martin}, \citenamefont {Martin}, \citenamefont {Martin}, \citenamefont {Maxfield}, \citenamefont {McClean}, \citenamefont {McEwen}, \citenamefont {Meeks}, \citenamefont {Megrant}, \citenamefont {Mi}, \citenamefont {Miao}, \citenamefont
  {Mieszala}, \citenamefont {Molavi}, \citenamefont {Molina}, \citenamefont {Montazeri}, \citenamefont {Morvan}, \citenamefont {Movassagh}, \citenamefont {Mruczkiewicz}, \citenamefont {Naaman}, \citenamefont {Neeley}, \citenamefont {Neill}, \citenamefont {Nersisyan}, \citenamefont {Neven}, \citenamefont {Newman}, \citenamefont {Ng}, \citenamefont {Nguyen}, \citenamefont {Nguyen}, \citenamefont {Ni}, \citenamefont {Niu}, \citenamefont {O'Brien}, \citenamefont {Oliver}, \citenamefont {Opremcak}, \citenamefont {Ottosson}, \citenamefont {Petukhov}, \citenamefont {Pizzuto}, \citenamefont {Platt}, \citenamefont {Potter}, \citenamefont {Pritchard}, \citenamefont {Pryadko}, \citenamefont {Quintana}, \citenamefont {Ramachandran}, \citenamefont {Reagor}, \citenamefont {Redding}, \citenamefont {Rhodes}, \citenamefont {Roberts}, \citenamefont {Rosenberg}, \citenamefont {Rosenfeld}, \citenamefont {Roushan}, \citenamefont {Rubin}, \citenamefont {Saei}, \citenamefont {Sank}, \citenamefont {Sankaragomathi}, \citenamefont
  {Satzinger}, \citenamefont {Schurkus}, \citenamefont {Schuster}, \citenamefont {Senior}, \citenamefont {Shearn}, \citenamefont {Shorter}, \citenamefont {Shutty}, \citenamefont {Shvarts}, \citenamefont {Singh}, \citenamefont {Sivak}, \citenamefont {Skruzny}, \citenamefont {Small}, \citenamefont {Smelyanskiy}, \citenamefont {Smith}, \citenamefont {Somma}, \citenamefont {Springer}, \citenamefont {Sterling}, \citenamefont {Strain}, \citenamefont {Suchard}, \citenamefont {Szasz}, \citenamefont {Sztein}, \citenamefont {Thor}, \citenamefont {Torres}, \citenamefont {Torunbalci}, \citenamefont {Vaishnav}, \citenamefont {Vargas}, \citenamefont {Vdovichev}, \citenamefont {Vidal}, \citenamefont {Villalonga}, \citenamefont {Heidweiller}, \citenamefont {Waltman}, \citenamefont {Wang}, \citenamefont {Ware}, \citenamefont {Weber}, \citenamefont {Weidel}, \citenamefont {White}, \citenamefont {Wong}, \citenamefont {Woo}, \citenamefont {Xing}, \citenamefont {Yao}, \citenamefont {Yeh}, \citenamefont {Ying}, \citenamefont
  {Yoo}, \citenamefont {Yosri}, \citenamefont {Young}, \citenamefont {Zalcman}, \citenamefont {Zhang}, \citenamefont {Zhu},\ and\ \citenamefont {Zobrist}}]{google2024belowthesurfacecode}%
  \BibitemOpen
  \bibfield  {author} {\bibinfo {author} {\bibfnamefont {R.}~\bibnamefont {Acharya}}, \bibinfo {author} {\bibfnamefont {D.~A.}\ \bibnamefont {Abanin}}, \bibinfo {author} {\bibfnamefont {L.}~\bibnamefont {Aghababaie-Beni}}, \bibinfo {author} {\bibfnamefont {I.}~\bibnamefont {Aleiner}}, \bibinfo {author} {\bibfnamefont {T.~I.}\ \bibnamefont {Andersen}}, \bibinfo {author} {\bibfnamefont {M.}~\bibnamefont {Ansmann}}, \bibinfo {author} {\bibfnamefont {F.}~\bibnamefont {Arute}}, \bibinfo {author} {\bibfnamefont {K.}~\bibnamefont {Arya}}, \bibinfo {author} {\bibfnamefont {A.}~\bibnamefont {Asfaw}}, \bibinfo {author} {\bibfnamefont {N.}~\bibnamefont {Astrakhantsev}}, \bibinfo {author} {\bibfnamefont {J.}~\bibnamefont {Atalaya}}, \bibinfo {author} {\bibfnamefont {R.}~\bibnamefont {Babbush}}, \bibinfo {author} {\bibfnamefont {D.}~\bibnamefont {Bacon}}, \bibinfo {author} {\bibfnamefont {B.}~\bibnamefont {Ballard}}, \bibinfo {author} {\bibfnamefont {J.~C.}\ \bibnamefont {Bardin}}, \bibinfo {author} {\bibfnamefont
  {J.}~\bibnamefont {Bausch}}, \bibinfo {author} {\bibfnamefont {A.}~\bibnamefont {Bengtsson}}, \bibinfo {author} {\bibfnamefont {A.}~\bibnamefont {Bilmes}}, \bibinfo {author} {\bibfnamefont {S.}~\bibnamefont {Blackwell}}, \bibinfo {author} {\bibfnamefont {S.}~\bibnamefont {Boixo}}, \bibinfo {author} {\bibfnamefont {G.}~\bibnamefont {Bortoli}}, \bibinfo {author} {\bibfnamefont {A.}~\bibnamefont {Bourassa}}, \bibinfo {author} {\bibfnamefont {J.}~\bibnamefont {Bovaird}}, \bibinfo {author} {\bibfnamefont {L.}~\bibnamefont {Brill}}, \bibinfo {author} {\bibfnamefont {M.}~\bibnamefont {Broughton}}, \bibinfo {author} {\bibfnamefont {D.~A.}\ \bibnamefont {Browne}}, \bibinfo {author} {\bibfnamefont {B.}~\bibnamefont {Buchea}}, \bibinfo {author} {\bibfnamefont {B.~B.}\ \bibnamefont {Buckley}}, \bibinfo {author} {\bibfnamefont {D.~A.}\ \bibnamefont {Buell}}, \bibinfo {author} {\bibfnamefont {T.}~\bibnamefont {Burger}}, \bibinfo {author} {\bibfnamefont {B.}~\bibnamefont {Burkett}}, \bibinfo {author} {\bibfnamefont
  {N.}~\bibnamefont {Bushnell}}, \bibinfo {author} {\bibfnamefont {A.}~\bibnamefont {Cabrera}}, \bibinfo {author} {\bibfnamefont {J.}~\bibnamefont {Campero}}, \bibinfo {author} {\bibfnamefont {H.-S.}\ \bibnamefont {Chang}}, \bibinfo {author} {\bibfnamefont {Y.}~\bibnamefont {Chen}}, \bibinfo {author} {\bibfnamefont {Z.}~\bibnamefont {Chen}}, \bibinfo {author} {\bibfnamefont {B.}~\bibnamefont {Chiaro}}, \bibinfo {author} {\bibfnamefont {D.}~\bibnamefont {Chik}}, \bibinfo {author} {\bibfnamefont {C.}~\bibnamefont {Chou}}, \bibinfo {author} {\bibfnamefont {J.}~\bibnamefont {Claes}}, \bibinfo {author} {\bibfnamefont {A.~Y.}\ \bibnamefont {Cleland}}, \bibinfo {author} {\bibfnamefont {J.}~\bibnamefont {Cogan}}, \bibinfo {author} {\bibfnamefont {R.}~\bibnamefont {Collins}}, \bibinfo {author} {\bibfnamefont {P.}~\bibnamefont {Conner}}, \bibinfo {author} {\bibfnamefont {W.}~\bibnamefont {Courtney}}, \bibinfo {author} {\bibfnamefont {A.~L.}\ \bibnamefont {Crook}}, \bibinfo {author} {\bibfnamefont {B.}~\bibnamefont
  {Curtin}}, \bibinfo {author} {\bibfnamefont {S.}~\bibnamefont {Das}}, \bibinfo {author} {\bibfnamefont {A.}~\bibnamefont {Davies}}, \bibinfo {author} {\bibfnamefont {L.}~\bibnamefont {De~Lorenzo}}, \bibinfo {author} {\bibfnamefont {D.~M.}\ \bibnamefont {Debroy}}, \bibinfo {author} {\bibfnamefont {S.}~\bibnamefont {Demura}}, \bibinfo {author} {\bibfnamefont {M.}~\bibnamefont {Devoret}}, \bibinfo {author} {\bibfnamefont {A.}~\bibnamefont {Di~Paolo}}, \bibinfo {author} {\bibfnamefont {P.}~\bibnamefont {Donohoe}}, \bibinfo {author} {\bibfnamefont {I.}~\bibnamefont {Drozdov}}, \bibinfo {author} {\bibfnamefont {A.}~\bibnamefont {Dunsworth}}, \bibinfo {author} {\bibfnamefont {C.}~\bibnamefont {Earle}}, \bibinfo {author} {\bibfnamefont {T.}~\bibnamefont {Edlich}}, \bibinfo {author} {\bibfnamefont {A.}~\bibnamefont {Eickbusch}}, \bibinfo {author} {\bibfnamefont {A.~M.}\ \bibnamefont {Elbag}}, \bibinfo {author} {\bibfnamefont {M.}~\bibnamefont {Elzouka}}, \bibinfo {author} {\bibfnamefont {C.}~\bibnamefont
  {Erickson}}, \bibinfo {author} {\bibfnamefont {L.}~\bibnamefont {Faoro}}, \bibinfo {author} {\bibfnamefont {E.}~\bibnamefont {Farhi}}, \bibinfo {author} {\bibfnamefont {V.~S.}\ \bibnamefont {Ferreira}}, \bibinfo {author} {\bibfnamefont {L.~F.}\ \bibnamefont {Burgos}}, \bibinfo {author} {\bibfnamefont {E.}~\bibnamefont {Forati}}, \bibinfo {author} {\bibfnamefont {A.~G.}\ \bibnamefont {Fowler}}, \bibinfo {author} {\bibfnamefont {B.}~\bibnamefont {Foxen}}, \bibinfo {author} {\bibfnamefont {S.}~\bibnamefont {Ganjam}}, \bibinfo {author} {\bibfnamefont {G.}~\bibnamefont {Garcia}}, \bibinfo {author} {\bibfnamefont {R.}~\bibnamefont {Gasca}}, \bibinfo {author} {\bibfnamefont {Ã.}~\bibnamefont {Genois}}, \bibinfo {author} {\bibfnamefont {W.}~\bibnamefont {Giang}}, \bibinfo {author} {\bibfnamefont {C.}~\bibnamefont {Gidney}}, \bibinfo {author} {\bibfnamefont {D.}~\bibnamefont {Gilboa}}, \bibinfo {author} {\bibfnamefont {R.}~\bibnamefont {Gosula}}, \bibinfo {author} {\bibfnamefont {A.~G.}\ \bibnamefont {Dau}},
  \bibinfo {author} {\bibfnamefont {D.}~\bibnamefont {Graumann}}, \bibinfo {author} {\bibfnamefont {A.}~\bibnamefont {Greene}}, \bibinfo {author} {\bibfnamefont {J.~A.}\ \bibnamefont {Gross}}, \bibinfo {author} {\bibfnamefont {S.}~\bibnamefont {Habegger}}, \bibinfo {author} {\bibfnamefont {J.}~\bibnamefont {Hall}}, \bibinfo {author} {\bibfnamefont {M.~C.}\ \bibnamefont {Hamilton}}, \bibinfo {author} {\bibfnamefont {M.}~\bibnamefont {Hansen}}, \bibinfo {author} {\bibfnamefont {M.~P.}\ \bibnamefont {Harrigan}}, \bibinfo {author} {\bibfnamefont {S.~D.}\ \bibnamefont {Harrington}}, \bibinfo {author} {\bibfnamefont {F.~J.~H.}\ \bibnamefont {Heras}}, \bibinfo {author} {\bibfnamefont {S.}~\bibnamefont {Heslin}}, \bibinfo {author} {\bibfnamefont {P.}~\bibnamefont {Heu}}, \bibinfo {author} {\bibfnamefont {O.}~\bibnamefont {Higgott}}, \bibinfo {author} {\bibfnamefont {G.}~\bibnamefont {Hill}}, \bibinfo {author} {\bibfnamefont {J.}~\bibnamefont {Hilton}}, \bibinfo {author} {\bibfnamefont {G.}~\bibnamefont {Holland}},
  \bibinfo {author} {\bibfnamefont {S.}~\bibnamefont {Hong}}, \bibinfo {author} {\bibfnamefont {H.-Y.}\ \bibnamefont {Huang}}, \bibinfo {author} {\bibfnamefont {A.}~\bibnamefont {Huff}}, \bibinfo {author} {\bibfnamefont {W.~J.}\ \bibnamefont {Huggins}}, \bibinfo {author} {\bibfnamefont {L.~B.}\ \bibnamefont {Ioffe}}, \bibinfo {author} {\bibfnamefont {S.~V.}\ \bibnamefont {Isakov}}, \bibinfo {author} {\bibfnamefont {J.}~\bibnamefont {Iveland}}, \bibinfo {author} {\bibfnamefont {E.}~\bibnamefont {Jeffrey}}, \bibinfo {author} {\bibfnamefont {Z.}~\bibnamefont {Jiang}}, \bibinfo {author} {\bibfnamefont {C.}~\bibnamefont {Jones}}, \bibinfo {author} {\bibfnamefont {S.}~\bibnamefont {Jordan}}, \bibinfo {author} {\bibfnamefont {C.}~\bibnamefont {Joshi}}, \bibinfo {author} {\bibfnamefont {P.}~\bibnamefont {Juhas}}, \bibinfo {author} {\bibfnamefont {D.}~\bibnamefont {Kafri}}, \bibinfo {author} {\bibfnamefont {H.}~\bibnamefont {Kang}}, \bibinfo {author} {\bibfnamefont {A.~H.}\ \bibnamefont {Karamlou}}, \bibinfo {author}
  {\bibfnamefont {K.}~\bibnamefont {Kechedzhi}}, \bibinfo {author} {\bibfnamefont {J.}~\bibnamefont {Kelly}}, \bibinfo {author} {\bibfnamefont {T.}~\bibnamefont {Khaire}}, \bibinfo {author} {\bibfnamefont {T.}~\bibnamefont {Khattar}}, \bibinfo {author} {\bibfnamefont {M.}~\bibnamefont {Khezri}}, \bibinfo {author} {\bibfnamefont {S.}~\bibnamefont {Kim}}, \bibinfo {author} {\bibfnamefont {P.~V.}\ \bibnamefont {Klimov}}, \bibinfo {author} {\bibfnamefont {A.~R.}\ \bibnamefont {Klots}}, \bibinfo {author} {\bibfnamefont {B.}~\bibnamefont {Kobrin}}, \bibinfo {author} {\bibfnamefont {P.}~\bibnamefont {Kohli}}, \bibinfo {author} {\bibfnamefont {A.~N.}\ \bibnamefont {Korotkov}}, \bibinfo {author} {\bibfnamefont {F.}~\bibnamefont {Kostritsa}}, \bibinfo {author} {\bibfnamefont {R.}~\bibnamefont {Kothari}}, \bibinfo {author} {\bibfnamefont {B.}~\bibnamefont {Kozlovskii}}, \bibinfo {author} {\bibfnamefont {J.~M.}\ \bibnamefont {Kreikebaum}}, \bibinfo {author} {\bibfnamefont {V.~D.}\ \bibnamefont {Kurilovich}}, \bibinfo
  {author} {\bibfnamefont {N.}~\bibnamefont {Lacroix}}, \bibinfo {author} {\bibfnamefont {D.}~\bibnamefont {Landhuis}}, \bibinfo {author} {\bibfnamefont {T.}~\bibnamefont {Lange-Dei}}, \bibinfo {author} {\bibfnamefont {B.~W.}\ \bibnamefont {Langley}}, \bibinfo {author} {\bibfnamefont {P.}~\bibnamefont {Laptev}}, \bibinfo {author} {\bibfnamefont {K.-M.}\ \bibnamefont {Lau}}, \bibinfo {author} {\bibfnamefont {L.}~\bibnamefont {Le~Guevel}}, \bibinfo {author} {\bibfnamefont {J.}~\bibnamefont {Ledford}}, \bibinfo {author} {\bibfnamefont {J.}~\bibnamefont {Lee}}, \bibinfo {author} {\bibfnamefont {K.}~\bibnamefont {Lee}}, \bibinfo {author} {\bibfnamefont {Y.~D.}\ \bibnamefont {Lensky}}, \bibinfo {author} {\bibfnamefont {S.}~\bibnamefont {Leon}}, \bibinfo {author} {\bibfnamefont {B.~J.}\ \bibnamefont {Lester}}, \bibinfo {author} {\bibfnamefont {W.~Y.}\ \bibnamefont {Li}}, \bibinfo {author} {\bibfnamefont {Y.}~\bibnamefont {Li}}, \bibinfo {author} {\bibfnamefont {A.~T.}\ \bibnamefont {Lill}}, \bibinfo {author}
  {\bibfnamefont {W.}~\bibnamefont {Liu}}, \bibinfo {author} {\bibfnamefont {W.~P.}\ \bibnamefont {Livingston}}, \bibinfo {author} {\bibfnamefont {A.}~\bibnamefont {Locharla}}, \bibinfo {author} {\bibfnamefont {E.}~\bibnamefont {Lucero}}, \bibinfo {author} {\bibfnamefont {D.}~\bibnamefont {Lundahl}}, \bibinfo {author} {\bibfnamefont {A.}~\bibnamefont {Lunt}}, \bibinfo {author} {\bibfnamefont {S.}~\bibnamefont {Madhuk}}, \bibinfo {author} {\bibfnamefont {F.~D.}\ \bibnamefont {Malone}}, \bibinfo {author} {\bibfnamefont {A.}~\bibnamefont {Maloney}}, \bibinfo {author} {\bibfnamefont {S.}~\bibnamefont {Mandrà}}, \bibinfo {author} {\bibfnamefont {J.}~\bibnamefont {Manyika}}, \bibinfo {author} {\bibfnamefont {L.~S.}\ \bibnamefont {Martin}}, \bibinfo {author} {\bibfnamefont {O.}~\bibnamefont {Martin}}, \bibinfo {author} {\bibfnamefont {S.}~\bibnamefont {Martin}}, \bibinfo {author} {\bibfnamefont {C.}~\bibnamefont {Maxfield}}, \bibinfo {author} {\bibfnamefont {J.~R.}\ \bibnamefont {McClean}}, \bibinfo {author}
  {\bibfnamefont {M.}~\bibnamefont {McEwen}}, \bibinfo {author} {\bibfnamefont {S.}~\bibnamefont {Meeks}}, \bibinfo {author} {\bibfnamefont {A.}~\bibnamefont {Megrant}}, \bibinfo {author} {\bibfnamefont {X.}~\bibnamefont {Mi}}, \bibinfo {author} {\bibfnamefont {K.~C.}\ \bibnamefont {Miao}}, \bibinfo {author} {\bibfnamefont {A.}~\bibnamefont {Mieszala}}, \bibinfo {author} {\bibfnamefont {R.}~\bibnamefont {Molavi}}, \bibinfo {author} {\bibfnamefont {S.}~\bibnamefont {Molina}}, \bibinfo {author} {\bibfnamefont {S.}~\bibnamefont {Montazeri}}, \bibinfo {author} {\bibfnamefont {A.}~\bibnamefont {Morvan}}, \bibinfo {author} {\bibfnamefont {R.}~\bibnamefont {Movassagh}}, \bibinfo {author} {\bibfnamefont {W.}~\bibnamefont {Mruczkiewicz}}, \bibinfo {author} {\bibfnamefont {O.}~\bibnamefont {Naaman}}, \bibinfo {author} {\bibfnamefont {M.}~\bibnamefont {Neeley}}, \bibinfo {author} {\bibfnamefont {C.}~\bibnamefont {Neill}}, \bibinfo {author} {\bibfnamefont {A.}~\bibnamefont {Nersisyan}}, \bibinfo {author} {\bibfnamefont
  {H.}~\bibnamefont {Neven}}, \bibinfo {author} {\bibfnamefont {M.}~\bibnamefont {Newman}}, \bibinfo {author} {\bibfnamefont {J.~H.}\ \bibnamefont {Ng}}, \bibinfo {author} {\bibfnamefont {A.}~\bibnamefont {Nguyen}}, \bibinfo {author} {\bibfnamefont {M.}~\bibnamefont {Nguyen}}, \bibinfo {author} {\bibfnamefont {C.-H.}\ \bibnamefont {Ni}}, \bibinfo {author} {\bibfnamefont {M.~Y.}\ \bibnamefont {Niu}}, \bibinfo {author} {\bibfnamefont {T.~E.}\ \bibnamefont {O'Brien}}, \bibinfo {author} {\bibfnamefont {W.~D.}\ \bibnamefont {Oliver}}, \bibinfo {author} {\bibfnamefont {A.}~\bibnamefont {Opremcak}}, \bibinfo {author} {\bibfnamefont {K.}~\bibnamefont {Ottosson}}, \bibinfo {author} {\bibfnamefont {A.}~\bibnamefont {Petukhov}}, \bibinfo {author} {\bibfnamefont {A.}~\bibnamefont {Pizzuto}}, \bibinfo {author} {\bibfnamefont {J.}~\bibnamefont {Platt}}, \bibinfo {author} {\bibfnamefont {R.}~\bibnamefont {Potter}}, \bibinfo {author} {\bibfnamefont {O.}~\bibnamefont {Pritchard}}, \bibinfo {author} {\bibfnamefont {L.~P.}\
  \bibnamefont {Pryadko}}, \bibinfo {author} {\bibfnamefont {C.}~\bibnamefont {Quintana}}, \bibinfo {author} {\bibfnamefont {G.}~\bibnamefont {Ramachandran}}, \bibinfo {author} {\bibfnamefont {M.~J.}\ \bibnamefont {Reagor}}, \bibinfo {author} {\bibfnamefont {J.}~\bibnamefont {Redding}}, \bibinfo {author} {\bibfnamefont {D.~M.}\ \bibnamefont {Rhodes}}, \bibinfo {author} {\bibfnamefont {G.}~\bibnamefont {Roberts}}, \bibinfo {author} {\bibfnamefont {E.}~\bibnamefont {Rosenberg}}, \bibinfo {author} {\bibfnamefont {E.}~\bibnamefont {Rosenfeld}}, \bibinfo {author} {\bibfnamefont {P.}~\bibnamefont {Roushan}}, \bibinfo {author} {\bibfnamefont {N.~C.}\ \bibnamefont {Rubin}}, \bibinfo {author} {\bibfnamefont {N.}~\bibnamefont {Saei}}, \bibinfo {author} {\bibfnamefont {D.}~\bibnamefont {Sank}}, \bibinfo {author} {\bibfnamefont {K.}~\bibnamefont {Sankaragomathi}}, \bibinfo {author} {\bibfnamefont {K.~J.}\ \bibnamefont {Satzinger}}, \bibinfo {author} {\bibfnamefont {H.~F.}\ \bibnamefont {Schurkus}}, \bibinfo {author}
  {\bibfnamefont {C.}~\bibnamefont {Schuster}}, \bibinfo {author} {\bibfnamefont {A.~W.}\ \bibnamefont {Senior}}, \bibinfo {author} {\bibfnamefont {M.~J.}\ \bibnamefont {Shearn}}, \bibinfo {author} {\bibfnamefont {A.}~\bibnamefont {Shorter}}, \bibinfo {author} {\bibfnamefont {N.}~\bibnamefont {Shutty}}, \bibinfo {author} {\bibfnamefont {V.}~\bibnamefont {Shvarts}}, \bibinfo {author} {\bibfnamefont {S.}~\bibnamefont {Singh}}, \bibinfo {author} {\bibfnamefont {V.}~\bibnamefont {Sivak}}, \bibinfo {author} {\bibfnamefont {J.}~\bibnamefont {Skruzny}}, \bibinfo {author} {\bibfnamefont {S.}~\bibnamefont {Small}}, \bibinfo {author} {\bibfnamefont {V.}~\bibnamefont {Smelyanskiy}}, \bibinfo {author} {\bibfnamefont {W.~C.}\ \bibnamefont {Smith}}, \bibinfo {author} {\bibfnamefont {R.~D.}\ \bibnamefont {Somma}}, \bibinfo {author} {\bibfnamefont {S.}~\bibnamefont {Springer}}, \bibinfo {author} {\bibfnamefont {G.}~\bibnamefont {Sterling}}, \bibinfo {author} {\bibfnamefont {D.}~\bibnamefont {Strain}}, \bibinfo {author}
  {\bibfnamefont {J.}~\bibnamefont {Suchard}}, \bibinfo {author} {\bibfnamefont {A.}~\bibnamefont {Szasz}}, \bibinfo {author} {\bibfnamefont {A.}~\bibnamefont {Sztein}}, \bibinfo {author} {\bibfnamefont {D.}~\bibnamefont {Thor}}, \bibinfo {author} {\bibfnamefont {A.}~\bibnamefont {Torres}}, \bibinfo {author} {\bibfnamefont {M.~M.}\ \bibnamefont {Torunbalci}}, \bibinfo {author} {\bibfnamefont {A.}~\bibnamefont {Vaishnav}}, \bibinfo {author} {\bibfnamefont {J.}~\bibnamefont {Vargas}}, \bibinfo {author} {\bibfnamefont {S.}~\bibnamefont {Vdovichev}}, \bibinfo {author} {\bibfnamefont {G.}~\bibnamefont {Vidal}}, \bibinfo {author} {\bibfnamefont {B.}~\bibnamefont {Villalonga}}, \bibinfo {author} {\bibfnamefont {C.~V.}\ \bibnamefont {Heidweiller}}, \bibinfo {author} {\bibfnamefont {S.}~\bibnamefont {Waltman}}, \bibinfo {author} {\bibfnamefont {S.~X.}\ \bibnamefont {Wang}}, \bibinfo {author} {\bibfnamefont {B.}~\bibnamefont {Ware}}, \bibinfo {author} {\bibfnamefont {K.}~\bibnamefont {Weber}}, \bibinfo {author}
  {\bibfnamefont {T.}~\bibnamefont {Weidel}}, \bibinfo {author} {\bibfnamefont {T.}~\bibnamefont {White}}, \bibinfo {author} {\bibfnamefont {K.}~\bibnamefont {Wong}}, \bibinfo {author} {\bibfnamefont {B.~W.~K.}\ \bibnamefont {Woo}}, \bibinfo {author} {\bibfnamefont {C.}~\bibnamefont {Xing}}, \bibinfo {author} {\bibfnamefont {Z.~J.}\ \bibnamefont {Yao}}, \bibinfo {author} {\bibfnamefont {P.}~\bibnamefont {Yeh}}, \bibinfo {author} {\bibfnamefont {B.}~\bibnamefont {Ying}}, \bibinfo {author} {\bibfnamefont {J.}~\bibnamefont {Yoo}}, \bibinfo {author} {\bibfnamefont {N.}~\bibnamefont {Yosri}}, \bibinfo {author} {\bibfnamefont {G.}~\bibnamefont {Young}}, \bibinfo {author} {\bibfnamefont {A.}~\bibnamefont {Zalcman}}, \bibinfo {author} {\bibfnamefont {Y.}~\bibnamefont {Zhang}}, \bibinfo {author} {\bibfnamefont {N.}~\bibnamefont {Zhu}},\ and\ \bibinfo {author} {\bibfnamefont {N.}~\bibnamefont {Zobrist}},\ }\bibfield  {title} {\bibinfo {title} {Quantum error correction below the surface code threshold},\ }\href
  {https://doi.org/10.1038/s41586-024-08449-y} {\bibfield  {journal} {\bibinfo  {journal} {Nature}\ }\textbf {\bibinfo {volume} {638}},\ \bibinfo {pages} {920–926} (\bibinfo {year} {2024})}\BibitemShut {NoStop}%
\bibitem [{\citenamefont {Gidney}\ and\ \citenamefont {Eker{\aa}}(2021)}]{gidney2021factor}%
  \BibitemOpen
  \bibfield  {author} {\bibinfo {author} {\bibfnamefont {C.}~\bibnamefont {Gidney}}\ and\ \bibinfo {author} {\bibfnamefont {M.}~\bibnamefont {Eker{\aa}}},\ }\bibfield  {title} {\bibinfo {title} {How to factor 2048 bit rsa integers in 8 hours using 20 million noisy qubits},\ }\href {https://doi.org/10.22331/q-2021-04-15-433} {\bibfield  {journal} {\bibinfo  {journal} {Quantum}\ }\textbf {\bibinfo {volume} {5}},\ \bibinfo {pages} {433} (\bibinfo {year} {2021})}\BibitemShut {NoStop}%
\bibitem [{\citenamefont {Yoshioka}\ \emph {et~al.}(2024)\citenamefont {Yoshioka}, \citenamefont {Okubo}, \citenamefont {Suzuki}, \citenamefont {Koizumi},\ and\ \citenamefont {Mizukami}}]{yoshioka2022hunting}%
  \BibitemOpen
  \bibfield  {author} {\bibinfo {author} {\bibfnamefont {N.}~\bibnamefont {Yoshioka}}, \bibinfo {author} {\bibfnamefont {T.}~\bibnamefont {Okubo}}, \bibinfo {author} {\bibfnamefont {Y.}~\bibnamefont {Suzuki}}, \bibinfo {author} {\bibfnamefont {Y.}~\bibnamefont {Koizumi}},\ and\ \bibinfo {author} {\bibfnamefont {W.}~\bibnamefont {Mizukami}},\ }\bibfield  {title} {\bibinfo {title} {Hunting for quantum-classical crossover in condensed matter problems},\ }\bibfield  {journal} {\bibinfo  {journal} {npj Quantum Information}\ }\textbf {\bibinfo {volume} {10}},\ \href {https://doi.org/10.1038/s41534-024-00839-4} {10.1038/s41534-024-00839-4} (\bibinfo {year} {2024})\BibitemShut {NoStop}%
\bibitem [{\citenamefont {Reiher}\ \emph {et~al.}(2017)\citenamefont {Reiher}, \citenamefont {Wiebe}, \citenamefont {Svore}, \citenamefont {Wecker},\ and\ \citenamefont {Troyer}}]{reiher2017elucidating}%
  \BibitemOpen
  \bibfield  {author} {\bibinfo {author} {\bibfnamefont {M.}~\bibnamefont {Reiher}}, \bibinfo {author} {\bibfnamefont {N.}~\bibnamefont {Wiebe}}, \bibinfo {author} {\bibfnamefont {K.~M.}\ \bibnamefont {Svore}}, \bibinfo {author} {\bibfnamefont {D.}~\bibnamefont {Wecker}},\ and\ \bibinfo {author} {\bibfnamefont {M.}~\bibnamefont {Troyer}},\ }\bibfield  {title} {\bibinfo {title} {Elucidating reaction mechanisms on quantum computers},\ }\href {https://doi.org/10.1073/pnas.1619152114} {\bibfield  {journal} {\bibinfo  {journal} {Proceedings of the national academy of sciences}\ }\textbf {\bibinfo {volume} {114}},\ \bibinfo {pages} {7555} (\bibinfo {year} {2017})}\BibitemShut {NoStop}%
\bibitem [{\citenamefont {Goings}\ \emph {et~al.}(2022)\citenamefont {Goings}, \citenamefont {White}, \citenamefont {Lee}, \citenamefont {Tautermann}, \citenamefont {Degroote}, \citenamefont {Gidney}, \citenamefont {Shiozaki}, \citenamefont {Babbush},\ and\ \citenamefont {Rubin}}]{goings2022reliably}%
  \BibitemOpen
  \bibfield  {author} {\bibinfo {author} {\bibfnamefont {J.~J.}\ \bibnamefont {Goings}}, \bibinfo {author} {\bibfnamefont {A.}~\bibnamefont {White}}, \bibinfo {author} {\bibfnamefont {J.}~\bibnamefont {Lee}}, \bibinfo {author} {\bibfnamefont {C.~S.}\ \bibnamefont {Tautermann}}, \bibinfo {author} {\bibfnamefont {M.}~\bibnamefont {Degroote}}, \bibinfo {author} {\bibfnamefont {C.}~\bibnamefont {Gidney}}, \bibinfo {author} {\bibfnamefont {T.}~\bibnamefont {Shiozaki}}, \bibinfo {author} {\bibfnamefont {R.}~\bibnamefont {Babbush}},\ and\ \bibinfo {author} {\bibfnamefont {N.~C.}\ \bibnamefont {Rubin}},\ }\bibfield  {title} {\bibinfo {title} {Reliably assessing the electronic structure of cytochrome p450 on today's classical computers and tomorrow's quantum computers},\ }\bibfield  {journal} {\bibinfo  {journal} {Proceedings of the National Academy of Sciences}\ }\textbf {\bibinfo {volume} {119}},\ \href {https://doi.org/10.1073/pnas.2203533119} {10.1073/pnas.2203533119} (\bibinfo {year} {2022})\BibitemShut
  {NoStop}%
\bibitem [{\citenamefont {Lin}\ and\ \citenamefont {Tong}(2022)}]{PRXQuantum.3.010318}%
  \BibitemOpen
  \bibfield  {author} {\bibinfo {author} {\bibfnamefont {L.}~\bibnamefont {Lin}}\ and\ \bibinfo {author} {\bibfnamefont {Y.}~\bibnamefont {Tong}},\ }\bibfield  {title} {\bibinfo {title} {Heisenberg-limited ground-state energy estimation for early fault-tolerant quantum computers},\ }\href {https://doi.org/10.1103/PRXQuantum.3.010318} {\bibfield  {journal} {\bibinfo  {journal} {PRX Quantum}\ }\textbf {\bibinfo {volume} {3}},\ \bibinfo {pages} {010318} (\bibinfo {year} {2022})}\BibitemShut {NoStop}%
\bibitem [{\citenamefont {Wan}\ \emph {et~al.}(2022)\citenamefont {Wan}, \citenamefont {Berta},\ and\ \citenamefont {Campbell}}]{PhysRevLett.129.030503}%
  \BibitemOpen
  \bibfield  {author} {\bibinfo {author} {\bibfnamefont {K.}~\bibnamefont {Wan}}, \bibinfo {author} {\bibfnamefont {M.}~\bibnamefont {Berta}},\ and\ \bibinfo {author} {\bibfnamefont {E.~T.}\ \bibnamefont {Campbell}},\ }\bibfield  {title} {\bibinfo {title} {Randomized quantum algorithm for statistical phase estimation},\ }\href {https://doi.org/10.1103/PhysRevLett.129.030503} {\bibfield  {journal} {\bibinfo  {journal} {Phys. Rev. Lett.}\ }\textbf {\bibinfo {volume} {129}},\ \bibinfo {pages} {030503} (\bibinfo {year} {2022})}\BibitemShut {NoStop}%
\bibitem [{\citenamefont {Kshirsagar}\ \emph {et~al.}(2022)\citenamefont {Kshirsagar}, \citenamefont {Katabarwa},\ and\ \citenamefont {Johnson}}]{https://doi.org/10.48550/arxiv.2209.11322}%
  \BibitemOpen
  \bibfield  {author} {\bibinfo {author} {\bibfnamefont {R.}~\bibnamefont {Kshirsagar}}, \bibinfo {author} {\bibfnamefont {A.}~\bibnamefont {Katabarwa}},\ and\ \bibinfo {author} {\bibfnamefont {P.~D.}\ \bibnamefont {Johnson}},\ }\href {https://doi.org/10.48550/ARXIV.2209.11322} {\bibinfo {title} {On proving the robustness of algorithms for early fault-tolerant quantum computers}} (\bibinfo {year} {2022}),\ \bibinfo {note} {https://arxiv.org/abs/2209.11322}\BibitemShut {NoStop}%
\bibitem [{\citenamefont {Ding}\ and\ \citenamefont {Lin}(2023)}]{PRXQuantum.4.020331}%
  \BibitemOpen
  \bibfield  {author} {\bibinfo {author} {\bibfnamefont {Z.}~\bibnamefont {Ding}}\ and\ \bibinfo {author} {\bibfnamefont {L.}~\bibnamefont {Lin}},\ }\bibfield  {title} {\bibinfo {title} {Even shorter quantum circuit for phase estimation on early fault-tolerant quantum computers with applications to ground-state energy estimation},\ }\href {https://doi.org/10.1103/PRXQuantum.4.020331} {\bibfield  {journal} {\bibinfo  {journal} {PRX Quantum}\ }\textbf {\bibinfo {volume} {4}},\ \bibinfo {pages} {020331} (\bibinfo {year} {2023})}\BibitemShut {NoStop}%
\bibitem [{\citenamefont {Suzuki}\ \emph {et~al.}(2022)\citenamefont {Suzuki}, \citenamefont {Endo}, \citenamefont {Fujii},\ and\ \citenamefont {Tokunaga}}]{suzuki2022quantum}%
  \BibitemOpen
  \bibfield  {author} {\bibinfo {author} {\bibfnamefont {Y.}~\bibnamefont {Suzuki}}, \bibinfo {author} {\bibfnamefont {S.}~\bibnamefont {Endo}}, \bibinfo {author} {\bibfnamefont {K.}~\bibnamefont {Fujii}},\ and\ \bibinfo {author} {\bibfnamefont {Y.}~\bibnamefont {Tokunaga}},\ }\bibfield  {title} {\bibinfo {title} {Quantum error mitigation as a universal error reduction technique: Applications from the nisq to the fault-tolerant quantum computing eras},\ }\href {https://doi.org/10.1103/PRXQuantum.3.010345} {\bibfield  {journal} {\bibinfo  {journal} {PRX Quantum}\ }\textbf {\bibinfo {volume} {3}},\ \bibinfo {pages} {010345} (\bibinfo {year} {2022})}\BibitemShut {NoStop}%
\bibitem [{\citenamefont {Piveteau}\ \emph {et~al.}(2021)\citenamefont {Piveteau}, \citenamefont {Sutter}, \citenamefont {Bravyi}, \citenamefont {Gambetta},\ and\ \citenamefont {Temme}}]{piveteau2021error}%
  \BibitemOpen
  \bibfield  {author} {\bibinfo {author} {\bibfnamefont {C.}~\bibnamefont {Piveteau}}, \bibinfo {author} {\bibfnamefont {D.}~\bibnamefont {Sutter}}, \bibinfo {author} {\bibfnamefont {S.}~\bibnamefont {Bravyi}}, \bibinfo {author} {\bibfnamefont {J.~M.}\ \bibnamefont {Gambetta}},\ and\ \bibinfo {author} {\bibfnamefont {K.}~\bibnamefont {Temme}},\ }\bibfield  {title} {\bibinfo {title} {Error mitigation for universal gates on encoded qubits},\ }\href {https://doi.org/10.1103/PhysRevLett.127.200505} {\bibfield  {journal} {\bibinfo  {journal} {Physical Review Letters}\ }\textbf {\bibinfo {volume} {127}},\ \bibinfo {pages} {200505} (\bibinfo {year} {2021})}\BibitemShut {NoStop}%
\bibitem [{\citenamefont {Breuckmann}\ and\ \citenamefont {Eberhardt}(2021)}]{PRXQuantum.2.040101}%
  \BibitemOpen
  \bibfield  {author} {\bibinfo {author} {\bibfnamefont {N.~P.}\ \bibnamefont {Breuckmann}}\ and\ \bibinfo {author} {\bibfnamefont {J.~N.}\ \bibnamefont {Eberhardt}},\ }\bibfield  {title} {\bibinfo {title} {Quantum low-density parity-check codes},\ }\href {https://doi.org/10.1103/PRXQuantum.2.040101} {\bibfield  {journal} {\bibinfo  {journal} {PRX Quantum}\ }\textbf {\bibinfo {volume} {2}},\ \bibinfo {pages} {040101} (\bibinfo {year} {2021})}\BibitemShut {NoStop}%
\bibitem [{\citenamefont {Panteleev}\ and\ \citenamefont {Kalachev}(2021)}]{Panteleev2021degeneratequantum}%
  \BibitemOpen
  \bibfield  {author} {\bibinfo {author} {\bibfnamefont {P.}~\bibnamefont {Panteleev}}\ and\ \bibinfo {author} {\bibfnamefont {G.}~\bibnamefont {Kalachev}},\ }\bibfield  {title} {\bibinfo {title} {Degenerate {Q}uantum {LDPC} {C}odes {W}ith {G}ood {F}inite {L}ength {P}erformance},\ }\href {https://doi.org/10.22331/q-2021-11-22-585} {\bibfield  {journal} {\bibinfo  {journal} {{Quantum}}\ }\textbf {\bibinfo {volume} {5}},\ \bibinfo {pages} {585} (\bibinfo {year} {2021})}\BibitemShut {NoStop}%
\bibitem [{\citenamefont {Gidney}\ \emph {et~al.}(2023)\citenamefont {Gidney}, \citenamefont {Newman}, \citenamefont {Brooks},\ and\ \citenamefont {Jones}}]{gidney2023yoked}%
  \BibitemOpen
  \bibfield  {author} {\bibinfo {author} {\bibfnamefont {C.}~\bibnamefont {Gidney}}, \bibinfo {author} {\bibfnamefont {M.}~\bibnamefont {Newman}}, \bibinfo {author} {\bibfnamefont {P.}~\bibnamefont {Brooks}},\ and\ \bibinfo {author} {\bibfnamefont {C.}~\bibnamefont {Jones}},\ }\href@noop {} {\bibinfo {title} {Yoked surface codes}} (\bibinfo {year} {2023}),\ \Eprint {https://arxiv.org/abs/2312.04522} {arXiv:2312.04522 [quant-ph]} \BibitemShut {NoStop}%
\bibitem [{\citenamefont {Hong}\ \emph {et~al.}(2024)\citenamefont {Hong}, \citenamefont {Marinelli}, \citenamefont {Kaufman},\ and\ \citenamefont {Lucas}}]{hong2024longrangeenhanced}%
  \BibitemOpen
  \bibfield  {author} {\bibinfo {author} {\bibfnamefont {Y.}~\bibnamefont {Hong}}, \bibinfo {author} {\bibfnamefont {M.}~\bibnamefont {Marinelli}}, \bibinfo {author} {\bibfnamefont {A.~M.}\ \bibnamefont {Kaufman}},\ and\ \bibinfo {author} {\bibfnamefont {A.}~\bibnamefont {Lucas}},\ }\href@noop {} {\bibinfo {title} {Long-range-enhanced surface codes}} (\bibinfo {year} {2024}),\ \Eprint {https://arxiv.org/abs/2309.11719} {arXiv:2309.11719 [quant-ph]} \BibitemShut {NoStop}%
\bibitem [{\citenamefont {Bravyi}\ \emph {et~al.}(2024)\citenamefont {Bravyi}, \citenamefont {Cross}, \citenamefont {Gambetta}, \citenamefont {Maslov}, \citenamefont {Rall},\ and\ \citenamefont {Yoder}}]{Bravyi_2024}%
  \BibitemOpen
  \bibfield  {author} {\bibinfo {author} {\bibfnamefont {S.}~\bibnamefont {Bravyi}}, \bibinfo {author} {\bibfnamefont {A.~W.}\ \bibnamefont {Cross}}, \bibinfo {author} {\bibfnamefont {J.~M.}\ \bibnamefont {Gambetta}}, \bibinfo {author} {\bibfnamefont {D.}~\bibnamefont {Maslov}}, \bibinfo {author} {\bibfnamefont {P.}~\bibnamefont {Rall}},\ and\ \bibinfo {author} {\bibfnamefont {T.~J.}\ \bibnamefont {Yoder}},\ }\bibfield  {title} {\bibinfo {title} {High-threshold and low-overhead fault-tolerant quantum memory},\ }\href {https://doi.org/10.1038/s41586-024-07107-7} {\bibfield  {journal} {\bibinfo  {journal} {Nature}\ }\textbf {\bibinfo {volume} {627}},\ \bibinfo {pages} {778–782} (\bibinfo {year} {2024})}\BibitemShut {NoStop}%
\bibitem [{\citenamefont {Goto}(2024)}]{goto2024manyhypercube}%
  \BibitemOpen
  \bibfield  {author} {\bibinfo {author} {\bibfnamefont {H.}~\bibnamefont {Goto}},\ }\href@noop {} {\bibinfo {title} {Many-hypercube codes: High-rate quantum error-correcting codes for high-performance fault-tolerant quantum computation}} (\bibinfo {year} {2024}),\ \Eprint {https://arxiv.org/abs/2403.16054} {arXiv:2403.16054 [quant-ph]} \BibitemShut {NoStop}%
\bibitem [{\citenamefont {Yoshida}\ \emph {et~al.}(2024)\citenamefont {Yoshida}, \citenamefont {Tamiya},\ and\ \citenamefont {Yamasaki}}]{yoshida2024concatenate}%
  \BibitemOpen
  \bibfield  {author} {\bibinfo {author} {\bibfnamefont {S.}~\bibnamefont {Yoshida}}, \bibinfo {author} {\bibfnamefont {S.}~\bibnamefont {Tamiya}},\ and\ \bibinfo {author} {\bibfnamefont {H.}~\bibnamefont {Yamasaki}},\ }\href@noop {} {\bibinfo {title} {Concatenate codes, save qubits}} (\bibinfo {year} {2024}),\ \Eprint {https://arxiv.org/abs/2402.09606} {arXiv:2402.09606 [quant-ph]} \BibitemShut {NoStop}%
\bibitem [{\citenamefont {Itogawa}\ \emph {et~al.}(2025)\citenamefont {Itogawa}, \citenamefont {Takada}, \citenamefont {Hirano},\ and\ \citenamefont {Fujii}}]{itogawa2025zeroleveldistill}%
  \BibitemOpen
  \bibfield  {author} {\bibinfo {author} {\bibfnamefont {T.}~\bibnamefont {Itogawa}}, \bibinfo {author} {\bibfnamefont {Y.}~\bibnamefont {Takada}}, \bibinfo {author} {\bibfnamefont {Y.}~\bibnamefont {Hirano}},\ and\ \bibinfo {author} {\bibfnamefont {K.}~\bibnamefont {Fujii}},\ }\bibfield  {title} {\bibinfo {title} {Efficient magic state distillation by zero-level distillation},\ }\href {https://doi.org/10.1103/thxx-njr6} {\bibfield  {journal} {\bibinfo  {journal} {PRX Quantum}\ }\textbf {\bibinfo {volume} {6}},\ \bibinfo {pages} {020356} (\bibinfo {year} {2025})}\BibitemShut {NoStop}%
\bibitem [{\citenamefont {Gidney}\ \emph {et~al.}(2024)\citenamefont {Gidney}, \citenamefont {Shutty},\ and\ \citenamefont {Jones}}]{gidney2024magicstatecultivationgrowing}%
  \BibitemOpen
  \bibfield  {author} {\bibinfo {author} {\bibfnamefont {C.}~\bibnamefont {Gidney}}, \bibinfo {author} {\bibfnamefont {N.}~\bibnamefont {Shutty}},\ and\ \bibinfo {author} {\bibfnamefont {C.}~\bibnamefont {Jones}},\ }\href {https://arxiv.org/abs/2409.17595} {\bibinfo {title} {Magic state cultivation: growing t states as cheap as cnot gates}} (\bibinfo {year} {2024}),\ \Eprint {https://arxiv.org/abs/2409.17595} {arXiv:2409.17595 [quant-ph]} \BibitemShut {NoStop}%
\bibitem [{\citenamefont {Akahoshi}\ \emph {et~al.}(2024{\natexlab{a}})\citenamefont {Akahoshi}, \citenamefont {Maruyama}, \citenamefont {Oshima}, \citenamefont {Sato},\ and\ \citenamefont {Fujii}}]{PRXQuantum.5.010337}%
  \BibitemOpen
  \bibfield  {author} {\bibinfo {author} {\bibfnamefont {Y.}~\bibnamefont {Akahoshi}}, \bibinfo {author} {\bibfnamefont {K.}~\bibnamefont {Maruyama}}, \bibinfo {author} {\bibfnamefont {H.}~\bibnamefont {Oshima}}, \bibinfo {author} {\bibfnamefont {S.}~\bibnamefont {Sato}},\ and\ \bibinfo {author} {\bibfnamefont {K.}~\bibnamefont {Fujii}},\ }\bibfield  {title} {\bibinfo {title} {Partially fault-tolerant quantum computing architecture with error-corrected clifford gates and space-time efficient analog rotations},\ }\href {https://doi.org/10.1103/PRXQuantum.5.010337} {\bibfield  {journal} {\bibinfo  {journal} {PRX Quantum}\ }\textbf {\bibinfo {volume} {5}},\ \bibinfo {pages} {010337} (\bibinfo {year} {2024}{\natexlab{a}})}\BibitemShut {NoStop}%
\bibitem [{\citenamefont {Toshio}\ \emph {et~al.}(2025)\citenamefont {Toshio}, \citenamefont {Akahoshi}, \citenamefont {Fujisaki}, \citenamefont {Oshima}, \citenamefont {Sato},\ and\ \citenamefont {Fujii}}]{PhysRevX.15.021057}%
  \BibitemOpen
  \bibfield  {author} {\bibinfo {author} {\bibfnamefont {R.}~\bibnamefont {Toshio}}, \bibinfo {author} {\bibfnamefont {Y.}~\bibnamefont {Akahoshi}}, \bibinfo {author} {\bibfnamefont {J.}~\bibnamefont {Fujisaki}}, \bibinfo {author} {\bibfnamefont {H.}~\bibnamefont {Oshima}}, \bibinfo {author} {\bibfnamefont {S.}~\bibnamefont {Sato}},\ and\ \bibinfo {author} {\bibfnamefont {K.}~\bibnamefont {Fujii}},\ }\bibfield  {title} {\bibinfo {title} {Practical quantum advantage on partially fault-tolerant quantum computer},\ }\href {https://doi.org/10.1103/PhysRevX.15.021057} {\bibfield  {journal} {\bibinfo  {journal} {Phys. Rev. X}\ }\textbf {\bibinfo {volume} {15}},\ \bibinfo {pages} {021057} (\bibinfo {year} {2025})}\BibitemShut {NoStop}%
\bibitem [{\citenamefont {Akahoshi}\ \emph {et~al.}(2024{\natexlab{b}})\citenamefont {Akahoshi}, \citenamefont {Toshio}, \citenamefont {Fujisaki}, \citenamefont {Oshima}, \citenamefont {Sato},\ and\ \citenamefont {Fujii}}]{akahoshi2024compilationtrotterbasedtimeevolution}%
  \BibitemOpen
  \bibfield  {author} {\bibinfo {author} {\bibfnamefont {Y.}~\bibnamefont {Akahoshi}}, \bibinfo {author} {\bibfnamefont {R.}~\bibnamefont {Toshio}}, \bibinfo {author} {\bibfnamefont {J.}~\bibnamefont {Fujisaki}}, \bibinfo {author} {\bibfnamefont {H.}~\bibnamefont {Oshima}}, \bibinfo {author} {\bibfnamefont {S.}~\bibnamefont {Sato}},\ and\ \bibinfo {author} {\bibfnamefont {K.}~\bibnamefont {Fujii}},\ }\href {https://arxiv.org/abs/2408.14929} {\bibinfo {title} {Compilation of trotter-based time evolution for partially fault-tolerant quantum computing architecture}} (\bibinfo {year} {2024}{\natexlab{b}}),\ \Eprint {https://arxiv.org/abs/2408.14929} {arXiv:2408.14929 [quant-ph]} \BibitemShut {NoStop}%
\bibitem [{\citenamefont {Dangwal}\ \emph {et~al.}(2025)\citenamefont {Dangwal}, \citenamefont {Vittal}, \citenamefont {Seifert}, \citenamefont {Chong},\ and\ \citenamefont {Ravi}}]{dangwal2025variationalquantumalgorithmsera}%
  \BibitemOpen
  \bibfield  {author} {\bibinfo {author} {\bibfnamefont {S.}~\bibnamefont {Dangwal}}, \bibinfo {author} {\bibfnamefont {S.}~\bibnamefont {Vittal}}, \bibinfo {author} {\bibfnamefont {L.~M.}\ \bibnamefont {Seifert}}, \bibinfo {author} {\bibfnamefont {F.~T.}\ \bibnamefont {Chong}},\ and\ \bibinfo {author} {\bibfnamefont {G.~S.}\ \bibnamefont {Ravi}},\ }\href {https://arxiv.org/abs/2503.20963} {\bibinfo {title} {Variational quantum algorithms in the era of early fault tolerance}} (\bibinfo {year} {2025}),\ \Eprint {https://arxiv.org/abs/2503.20963} {arXiv:2503.20963 [quant-ph]} \BibitemShut {NoStop}%
\bibitem [{\citenamefont {Hutter}\ \emph {et~al.}(2014)\citenamefont {Hutter}, \citenamefont {Wootton},\ and\ \citenamefont {Loss}}]{PhysRevA.89.022326}%
  \BibitemOpen
  \bibfield  {author} {\bibinfo {author} {\bibfnamefont {A.}~\bibnamefont {Hutter}}, \bibinfo {author} {\bibfnamefont {J.~R.}\ \bibnamefont {Wootton}},\ and\ \bibinfo {author} {\bibfnamefont {D.}~\bibnamefont {Loss}},\ }\bibfield  {title} {\bibinfo {title} {Efficient markov chain monte carlo algorithm for the surface code},\ }\href {https://doi.org/10.1103/PhysRevA.89.022326} {\bibfield  {journal} {\bibinfo  {journal} {Phys. Rev. A}\ }\textbf {\bibinfo {volume} {89}},\ \bibinfo {pages} {022326} (\bibinfo {year} {2014})}\BibitemShut {NoStop}%
\bibitem [{\citenamefont {Bomb\'{\i}n}\ \emph {et~al.}(2024)\citenamefont {Bomb\'{\i}n}, \citenamefont {Pant}, \citenamefont {Roberts},\ and\ \citenamefont {Seetharam}}]{PRXQuantum.5.010302}%
  \BibitemOpen
  \bibfield  {author} {\bibinfo {author} {\bibfnamefont {H.}~\bibnamefont {Bomb\'{\i}n}}, \bibinfo {author} {\bibfnamefont {M.}~\bibnamefont {Pant}}, \bibinfo {author} {\bibfnamefont {S.}~\bibnamefont {Roberts}},\ and\ \bibinfo {author} {\bibfnamefont {K.~I.}\ \bibnamefont {Seetharam}},\ }\bibfield  {title} {\bibinfo {title} {Fault-tolerant postselection for low-overhead magic state preparation},\ }\href {https://doi.org/10.1103/PRXQuantum.5.010302} {\bibfield  {journal} {\bibinfo  {journal} {PRX Quantum}\ }\textbf {\bibinfo {volume} {5}},\ \bibinfo {pages} {010302} (\bibinfo {year} {2024})}\BibitemShut {NoStop}%
\bibitem [{\citenamefont {Meister}\ \emph {et~al.}(2024)\citenamefont {Meister}, \citenamefont {Pattison},\ and\ \citenamefont {Preskill}}]{meister2024efficientsoftoutputdecoderssurface}%
  \BibitemOpen
  \bibfield  {author} {\bibinfo {author} {\bibfnamefont {N.}~\bibnamefont {Meister}}, \bibinfo {author} {\bibfnamefont {C.~A.}\ \bibnamefont {Pattison}},\ and\ \bibinfo {author} {\bibfnamefont {J.}~\bibnamefont {Preskill}},\ }\href {https://arxiv.org/abs/2405.07433} {\bibinfo {title} {Efficient soft-output decoders for the surface code}} (\bibinfo {year} {2024}),\ \Eprint {https://arxiv.org/abs/2405.07433} {arXiv:2405.07433 [quant-ph]} \BibitemShut {NoStop}%
\bibitem [{\citenamefont {Kishi}\ \emph {et~al.}()\citenamefont {Kishi}, \citenamefont {Toshio}, \citenamefont {Fujisaki}, \citenamefont {Oshima}, \citenamefont {Sato},\ and\ \citenamefont {Fujii}}]{kishi2025toappear}%
  \BibitemOpen
  \bibfield  {author} {\bibinfo {author} {\bibfnamefont {K.}~\bibnamefont {Kishi}}, \bibinfo {author} {\bibfnamefont {R.}~\bibnamefont {Toshio}}, \bibinfo {author} {\bibfnamefont {J.}~\bibnamefont {Fujisaki}}, \bibinfo {author} {\bibfnamefont {H.}~\bibnamefont {Oshima}}, \bibinfo {author} {\bibfnamefont {S.}~\bibnamefont {Sato}},\ and\ \bibinfo {author} {\bibfnamefont {K.}~\bibnamefont {Fujii}},\ }\bibfield  {title} {\bibinfo {title} {Early stopping for fast soft-output calculation in cluster-based decoders},\ }\href@noop {} {\bibinfo  {journal} {To appear}\ }\BibitemShut {NoStop}%
\bibitem [{\citenamefont {Smith}\ \emph {et~al.}(2024)\citenamefont {Smith}, \citenamefont {Brown},\ and\ \citenamefont {Bartlett}}]{Smith_2024}%
  \BibitemOpen
\bibfield  {journal} {  }\bibfield  {author} {\bibinfo {author} {\bibfnamefont {S.~C.}\ \bibnamefont {Smith}}, \bibinfo {author} {\bibfnamefont {B.~J.}\ \bibnamefont {Brown}},\ and\ \bibinfo {author} {\bibfnamefont {S.~D.}\ \bibnamefont {Bartlett}},\ }\bibfield  {title} {\bibinfo {title} {Mitigating errors in logical qubits},\ }\bibfield  {journal} {\bibinfo  {journal} {Communications Physics}\ }\textbf {\bibinfo {volume} {7}},\ \href {https://doi.org/10.1038/s42005-024-01883-4} {10.1038/s42005-024-01883-4} (\bibinfo {year} {2024})\BibitemShut {NoStop}%
\bibitem [{Note1()}]{Note1}%
  \BibitemOpen
  \bibinfo {note} {On preparation of the manuscript, we found a similar work~\cite {noah2025toappear}. Please see the note added in the last of Sec.~\ref {sec:summary}.}\BibitemShut {Stop}%
\bibitem [{\citenamefont {Chamberland}\ and\ \citenamefont {Campbell}(2022)}]{PRXQuantum.3.010331}%
  \BibitemOpen
  \bibfield  {author} {\bibinfo {author} {\bibfnamefont {C.}~\bibnamefont {Chamberland}}\ and\ \bibinfo {author} {\bibfnamefont {E.~T.}\ \bibnamefont {Campbell}},\ }\bibfield  {title} {\bibinfo {title} {Universal quantum computing with twist-free and temporally encoded lattice surgery},\ }\href {https://doi.org/10.1103/PRXQuantum.3.010331} {\bibfield  {journal} {\bibinfo  {journal} {PRX Quantum}\ }\textbf {\bibinfo {volume} {3}},\ \bibinfo {pages} {010331} (\bibinfo {year} {2022})}\BibitemShut {NoStop}%
\bibitem [{\citenamefont {Litinski}(2019)}]{Litinski2019gameofsurfacecodes}%
  \BibitemOpen
  \bibfield  {author} {\bibinfo {author} {\bibfnamefont {D.}~\bibnamefont {Litinski}},\ }\bibfield  {title} {\bibinfo {title} {A {G}ame of {S}urface {C}odes: {L}arge-{S}cale {Q}uantum {C}omputing with {L}attice {S}urgery},\ }\href {https://doi.org/10.22331/q-2019-03-05-128} {\bibfield  {journal} {\bibinfo  {journal} {{Quantum}}\ }\textbf {\bibinfo {volume} {3}},\ \bibinfo {pages} {128} (\bibinfo {year} {2019})}\BibitemShut {NoStop}%
\bibitem [{\citenamefont {Gidney}(2021)}]{Gidney_2021}%
  \BibitemOpen
  \bibfield  {author} {\bibinfo {author} {\bibfnamefont {C.}~\bibnamefont {Gidney}},\ }\bibfield  {title} {\bibinfo {title} {Stim: a fast stabilizer circuit simulator},\ }\href {https://doi.org/10.22331/q-2021-07-06-497} {\bibfield  {journal} {\bibinfo  {journal} {Quantum}\ }\textbf {\bibinfo {volume} {5}},\ \bibinfo {pages} {497} (\bibinfo {year} {2021})}\BibitemShut {NoStop}%
\bibitem [{\citenamefont {Higgott}\ and\ \citenamefont {Gidney}(2022)}]{pymatchingv2}%
  \BibitemOpen
  \bibfield  {author} {\bibinfo {author} {\bibfnamefont {O.}~\bibnamefont {Higgott}}\ and\ \bibinfo {author} {\bibfnamefont {C.}~\bibnamefont {Gidney}},\ }\href@noop {} {\bibinfo {title} {Pymatching v2}},\ \bibinfo {howpublished} {\url{https://github.com/oscarhiggott/PyMatching}} (\bibinfo {year} {2022})\BibitemShut {NoStop}%
\bibitem [{\citenamefont {Gidney}(2022)}]{Gidney2022stability}%
  \BibitemOpen
  \bibfield  {author} {\bibinfo {author} {\bibfnamefont {C.}~\bibnamefont {Gidney}},\ }\bibfield  {title} {\bibinfo {title} {Stability {E}xperiments: {T}he {O}verlooked {D}ual of {M}emory {E}xperiments},\ }\href {https://doi.org/10.22331/q-2022-08-24-786} {\bibfield  {journal} {\bibinfo  {journal} {{Quantum}}\ }\textbf {\bibinfo {volume} {6}},\ \bibinfo {pages} {786} (\bibinfo {year} {2022})}\BibitemShut {NoStop}%
\bibitem [{\citenamefont {Poulin}(2006)}]{PhysRevA.74.052333}%
  \BibitemOpen
  \bibfield  {author} {\bibinfo {author} {\bibfnamefont {D.}~\bibnamefont {Poulin}},\ }\bibfield  {title} {\bibinfo {title} {Optimal and efficient decoding of concatenated quantum block codes},\ }\href {https://doi.org/10.1103/PhysRevA.74.052333} {\bibfield  {journal} {\bibinfo  {journal} {Phys. Rev. A}\ }\textbf {\bibinfo {volume} {74}},\ \bibinfo {pages} {052333} (\bibinfo {year} {2006})}\BibitemShut {NoStop}%
\bibitem [{\citenamefont {Beverland}\ \emph {et~al.}(2022)\citenamefont {Beverland}, \citenamefont {Kliuchnikov},\ and\ \citenamefont {Schoute}}]{PRXQuantum.3.020342}%
  \BibitemOpen
  \bibfield  {author} {\bibinfo {author} {\bibfnamefont {M.}~\bibnamefont {Beverland}}, \bibinfo {author} {\bibfnamefont {V.}~\bibnamefont {Kliuchnikov}},\ and\ \bibinfo {author} {\bibfnamefont {E.}~\bibnamefont {Schoute}},\ }\bibfield  {title} {\bibinfo {title} {Surface code compilation via edge-disjoint paths},\ }\href {https://doi.org/10.1103/PRXQuantum.3.020342} {\bibfield  {journal} {\bibinfo  {journal} {PRX Quantum}\ }\textbf {\bibinfo {volume} {3}},\ \bibinfo {pages} {020342} (\bibinfo {year} {2022})}\BibitemShut {NoStop}%
\bibitem [{\citenamefont {Hamada}\ \emph {et~al.}(2024)\citenamefont {Hamada}, \citenamefont {Suzuki},\ and\ \citenamefont {Tokunaga}}]{hamada2024efficient}%
  \BibitemOpen
  \bibfield  {author} {\bibinfo {author} {\bibfnamefont {K.}~\bibnamefont {Hamada}}, \bibinfo {author} {\bibfnamefont {Y.}~\bibnamefont {Suzuki}},\ and\ \bibinfo {author} {\bibfnamefont {Y.}~\bibnamefont {Tokunaga}},\ }\href@noop {} {\bibinfo {title} {Efficient and high-performance routing of lattice-surgery paths on three-dimensional lattice}} (\bibinfo {year} {2024}),\ \Eprint {https://arxiv.org/abs/2401.15829} {arXiv:2401.15829 [quant-ph]} \BibitemShut {NoStop}%
\bibitem [{\citenamefont {Tan}\ \emph {et~al.}(2024)\citenamefont {Tan}, \citenamefont {Niu},\ and\ \citenamefont {Gidney}}]{tan2024sat}%
  \BibitemOpen
  \bibfield  {author} {\bibinfo {author} {\bibfnamefont {D.~B.}\ \bibnamefont {Tan}}, \bibinfo {author} {\bibfnamefont {M.~Y.}\ \bibnamefont {Niu}},\ and\ \bibinfo {author} {\bibfnamefont {C.}~\bibnamefont {Gidney}},\ }\href@noop {} {\bibinfo {title} {A sat scalpel for lattice surgery: Representation and synthesis of subroutines for surface-code fault-tolerant quantum computing}} (\bibinfo {year} {2024}),\ \Eprint {https://arxiv.org/abs/2404.18369} {arXiv:2404.18369 [quant-ph]} \BibitemShut {NoStop}%
\bibitem [{\citenamefont {Shutty}\ \emph {et~al.}()\citenamefont {Shutty}, \citenamefont {Gidney},\ and\ \citenamefont {Higgott}}]{noah2025toappear}%
  \BibitemOpen
  \bibfield  {author} {\bibinfo {author} {\bibfnamefont {N.}~\bibnamefont {Shutty}}, \bibinfo {author} {\bibfnamefont {C.}~\bibnamefont {Gidney}},\ and\ \bibinfo {author} {\bibfnamefont {O.}~\bibnamefont {Higgott}},\ }\bibfield  {title} {\bibinfo {title} {Early-stop lattice surgery},\ }\href {https://yale.hosted.panopto.com/Panopto/Pages/Viewer.aspx? id=dbfe6994-f408-46e5-8227-b33001046e13} {\bibinfo  {journal} {To appear}\ }\BibitemShut {NoStop}%
\end{thebibliography}%

\end{document}